\documentclass{emulateapj}
\usepackage{graphicx}
\usepackage{natbib}

\slugcomment{Accepted by ApJS, draft from \today}

\shorttitle{Hot high-mass accretion disk candidates}
\shortauthors{Beuther, Walsh \& Longmore}

\begin{document}


\title{Hot high-mass accretion disk candidates}


\author{H.~Beuther\altaffilmark{1}, A.J.~Walsh\altaffilmark{2}, S.N.~Longmore\altaffilmark{3}}





\altaffiltext{1}{Max-Planck-Institute for Astronomy, K\"onigstuhl 17,
              69117 Heidelberg, Germany, beuther@mpia.de}
\altaffiltext{2}{Centre for Astronomy, 
              School of Engineering and Physical Sciences,
              James Cook University,
              Townsville, QLD, 4811 Australia}
\altaffiltext{3}{Harvard-Smithsonian Center for Astrophysics,
              60 Garden Street, Cambridge, MA02138, USA}


\begin{abstract}
  To better understand the physical properties of accretion disks in
  high-mass star formation, we present a study of a dozen high-mass
  accretion disk candidates observed at high spatial resolution with
  the Australia Telescope Compact Array (ATCA) in the high-excitation
  (4,4) and (5,5) lines of NH$_3$. All of our originally selected
  sources were detected in both NH$_3$ transitions, directly
  associated with CH$_3$OH Class II maser emission and implying that
  high-excitation NH$_3$ lines are good tracers of the dense gas
  components in hot-core type targets. Only the one source that did
  not satisfy the initial selection criteria remained undetected. From
  the eleven mapped sources, six show clear signatures of rotation
  and/or infall motions.  These signatures vary from velocity
  gradients perpendicular to the outflows, to infall signatures in
  absorption against ultracompact H{\sc ii} regions, to more spherical
  infall signatures in emission. Although our spatial resolution is
  $\sim$1000\,AU, we do not find clear Keplerian signatures in any of
  the sources.  Furthermore, we also do not find flattened structures.
  In contrast to this, in several of the sources with rotational
  signatures, the spatial structure is approximately spherical with
  sizes exceeding $10^4$\,AU, showing considerable clumpy
  sub-structure at even smaller scales. This implies that on average
  typical Keplerian accretion disks -- if they exist as expected --
  should be confined to regions usually smaller than 1000\,AU. It is
  likely that these disks are fed by the larger-scale rotating
  envelope structure we observe here.  Furthermore, we do detect
  1.25\,cm continuum emission in most fields of view.  While in some
  cases weak cm continuum emission is associated with our targets,
  more typically larger-scale H{\sc ii} regions are seen offset more
  than $10''$ from our sources. While these H{\sc ii} regions are
  unlikely to be directly related to the target regions, this spatial
  association nevertheless additionally stresses that high-mass star
  formation rarely proceeds in an isolated fashion but in a clustered
  mode.
\end{abstract}




\section{Introduction}
\label{intro}

The characterization of accretion disks around young high-mass
protostars is one of the main unsolved questions in high-mass star
formation research \citep{beuther2006b,cesaroni2006,zinnecker2007}.
The controversy arises around the difficulty to accumulate mass onto a
high-mass protostar when it gets larger than 8\,M$_{\odot}$ because the
radiation pressure of the growing protostar may be strong enough to
revert the gas inflow (e.g., \citealt{kahn1974,wolfire1987}).
Different ways to circumvent this problem are proposed, the main are
disk accretion from a turbulent gas and dust core (e.g.,
\citealt{jijina1996,yorke2002,mckee2003}), competitive accretion and
potential (proto)stellar mergers at the dense centers of evolving
high-mass (proto)clusters (e.g.,
\citealt{bonnell2004,bonnell2006,bally2005}), and ionized accretion
flows continuing through the hypercompact H{\sc ii} region phase
(e.g., \citealt{keto2003,keto2007}).

Over recent years, much indirect evidence has been accumulated that
high-mass accretion disks do exist. The main argument stems from high-mass
molecular outflow observations that identify collimated and energetic
outflows from high-mass protostars, resembling the properties of known
low-mass star formation sites (e.g.,
\citealt{beuther2002d,beuther2005b,arce2006}).  Such collimated
jet-like outflow structures are only explainable if one assumes an
underlying high-mass accretion disk that drives these outflows via
magneto-centrifugal acceleration. From a theoretical modeling
approach, recent 2D and 3D magneto-hydrodynamical simulations of
high-mass collapsing gas cores result in the formation of high-mass
accretion disks as well \citep{yorke2002,krumholz2006b,kratter2006}.
Although alternative formation scenarios are proposed, there appears
to be a growing concensus in the high-mass star formation community that
accretion disks should also exist in high-mass star formation.
However, it is still poorly known whether such high-mass disks are
similar to their low-mass counterparts, hence dominated by the central
protostar and in Keplerian rotation, or whether they are perhaps
self-gravitating non-Keplerian entities.

While the indirect evidence for high-mass accretion disks is steadily
increasing, direct observational studies are largely missing. This
lack of observational evidence can be attributed to two main reasons.
The first is the clustered mode of high-mass star formation and the
typically large distances, hence spatially resolving and disentangling
such structures is a difficult task. The second difficulty is to
choose the right spectral line tracer which allows unambiguous
identification and characterization of the disk structure.  Many
spectral lines are either optically thick (e.g., CO, HCO$^+$, CS),
chemically difficult to interpret (e.g., HC$_3$N, HNCO) or excited in
the envelope and disk which causes confusion (e.g., HCN, CH$_3$CN).
To overcome these problems, we used the Australia Telescope Compact
Array (ATCA) at 1.2\,cm wavelengths including the most extended baselines
(resulting in a spatial resolution $\leq 1''$), and we aimed at the
highly excited NH$_3$(J,K) inversion lines (4,4) and (5,5).  NH$_3$ is
known to be a dense core tracer (e.g., \citealt{zhang1998a}), and the
high (J,K) inversion lines with excitation temperatures
($E_{\rm{lower}}$) of 200 and 295\,K, respectively, should only be
excited in the innermost and warm region close to the central
protostars. Similarly, \citet{osorio2009} also modeled the NH$_3$(4,4)
emission of the collapsing hot core G31.41.  Furthermore, radiation
transfer calculations for 3D hydro-simulations revealed that the
1.2\,cm band regime should be particularly well suited for such
studies because the inner disk regions may be optically thick at
frequencies above about 100\,GHz \citep{krumholz2007a}. This may make
future high-mass disk studies of the innermost regions at shorter
wavelengths with ALMA difficult.  Therefore, studying these lines at
high angular resolution with the ATCA allows us to penetrate deeply
into the natal cores and study the physical properties of the
predicted high-mass accretion disks.

Over the last few years there have been a lot of ``trial and error''
approaches for high-mass disk studies, but no consistent investigation
of a larger sample is public so far. Since the above outlined approach
has been proven very successful in the recent ATCA high-(J,K) NH$_3$
study of IRAS\,18089-1732 \citep{beuther2008a}, we are now aiming at a
source sample of twelve promising disk candidates mainly identified by
previous lower resolution NH$_3$ studies of 41 sources using the ATCA
\citep{longmore2007} and 60 sources using Mopra (Walsh et
al.~priv.~comm.). The proposed sources comprise the best
high-mass-disk-candidate sample for the southern hemisphere to date
(Sec.~\ref{sample}).

\section{Observations}

Our sample of twelve sources (see Table \ref{tablesample}) was
observed during six consecutive nights between July 1st and 7th 2008.
We always observed two sources per night in a track-sharing mode
cycling between the gain calibrators and sources. Table
\ref{tablesample} lists the corresponding gain calibrators (phase and
amplitude) for each pair of sources. Simultaneously, we observed the
NH$_3$(4,4) and (5,5) inversion lines with the frequencies of the main
hyperfine components at 24.139 and 24.533\,GHz, respectively. The
phase reference centers and velocities relative to the local standard
of rest ($v_{\rm{lsr}}$) are given in Table \ref{tablesample}.
Bandpass and flux were calibrated with observations of 1253-055 and
1934-638.  The spectral resolution of the observations was 62\,kHz
corresponding to a velocity resolution of $\sim 0.8$\,km\,s$^{-1}$.
The observations were conducted in the 1.5D configuration including
antenna 6 which resulted in a maximum baseline length of 4.3\,km.
Depending on the source structure (e.g., compact versus extended) and
strength we applied different weightings (robust values between -2 and
2) for each source and line. To better trace faint and/or extended
features, we occasionally excluded antenna 6 -- and hence the longest
baselines~-- from the imaging process (Table \ref{tablesample2}).  The
synthesized beams and rms are given in Table \ref{tablesample2}.

\begin{table*}[htb]
\footnotesize{
\caption{Observational parameters I}
\begin{tabular}{lrrrrrrr}
\hline \hline
Source & R.A. & Dec. & $v_{\rm{lsr}}$ & $d^b$ & cal & comment & Refs.\\
       & J200.0 & J2000.0 & $\frac{\rm{km}}{\rm{s}}$ & kpc & & & \\
\hline
G305.21+0.21 & 13:11:13.77 & -62:34:41.2 & -38.3 & 3.5 & 1352-63 & lm NE-SW, H$_2$ NE-SW & 1,2,15 \\
G316.81-0.06 & 14:45:26.90 & -59:49:16.3 & -38.7 & 2.7 & & lm N-S, No H$_2$,GF NNW-SSE,  & 2,4,5,13 \\
&&&&&& 7MM NNW-SSE? & \\
\hline
G323.74-0.26 & 15:31:45.80 & -56:30:49.9 & -49.6 & 3.3 & 1613-586 & lm SW-NE or E-W?, H$_2$ E-W?, GF ? & 1,4,5,6 \\
G327.3-0.60 & 15:53:09.29 & -54:36:57.5 & -46.0$^a$ & 3.1/11.2 &&  & 7 \\
\hline
G328.81+0.63 & 15:55:48.44 & -52:43:06.0 & -41.5 & 3.0 & 1613-586 & lm E-W?, H$_2$ E-W, SiO E-W & 1,2,3,4 \\
G331.28-0.19 & 16:11:26.90 & -51:41:56.6 & -88.1 & 5.4 &          & lm NNW-SSE or WNW-ESE?,  H$_2$ E-W & 1,2,3 \\
&&&&&& GF NE-SW, SiO NNE-SSW, 7mm E-W? & 4,5,13\\
\hline
G336.02-0.83 & 16:35:09.30 & -48:46:47.0 & -48.5 & 3.6/12.0 & 1600-44 & lm N-S & 4\\
G345.00-0.22 & 17:05:10.90 & -41:29:06.6 & -26.8$^a$ & 2.9/13.5 & & lm E-W? & 4\\
\hline
G351.77-0.54 & 17:26:42.57 & -36:09:17.6 & 1.2 & 2.2 & 1714-336 & lm NE-SW or N-S?, CO NE-SW & 1,4,8 \\
             &             &             &     &     &          & OH N-S, H$_2$O ring or NE-SW? & 16,17,18 \\
G0.55-0.85 & 17:50:14.53 & -28:54:30.7 & 18.0$^a$ & 7.7/9.4 & & & 4 \\
\hline
G19.47-0.17 & 18:25:54.70 & -11:52:34.1 & 19.7 & 1.9 & 1829-106 & CO NNE-SSW, GF ? & 5,13 \\
I18151-1208 & 18:17:58.24 & -12:07:24.5 & 32.8 & 3.0 & & H$2$/CO NW-SE, dust NE-SW, CH$_3$OH & 8,9,10,11,14 \\
\hline
\end{tabular}
~\\
{\footnotesize
$^a$ $v_{\rm{lsr}}$ taken from Mopra spectra (Walsh et al.~priv.~comm.).\\
$^b$ Most distances were taken from the literature. For those where we did not find distance references, we calculated the kinematic near and far distances using the Galactic rotation curve by \citet{brand1993}.\\
  Comments: lm $\rightarrow$ linear maser with orientation, H$_2$/SiO/CO/7mm outflows with potential orientation, GF $\rightarrow$ ``Green fuzzy'' Spitzer 4.5\,$\mu$m elongation, dust continuum orientation, CH$_3$OH for the last source just states Class II CH$_3$OH maser detection. ``?'' denotes uncertainty.\\
  (1) \citet{norris1993}, (2) \citet{debuizer2003}, (3) \citet{debuizer2008}, (4) \citet{walsh1998}, (5) \citet{longmore2007}, (6) \citet{walsh2002}, (7) \citet{caswell1995}, (8) \citet{leurini2008}, (9) \citet{sridha}, (10) \citet{beuther2002b}, (11) \citet{davis2004}, (12) Fallscheer et al.~(in prep.), (13) \citet{longmore2009}, (14) \citet{longmore2009b}, (15) \citet{beuther2002c}, (16) \citet{longmore2007b}, (17) \citet{fish2005}, (18) \citet{forster1990}, (19) \citet{zapata2008}, (20) Longmore et al.~(in prep.)}
\label{tablesample}
}
\end{table*}

\begin{table}[htb]
\caption{Observational parameters II}
\begin{tabular}{lrrrr}
\hline \hline
Source & Line/Cont. & Beam & $1\sigma$\,rms$^a$ & Peak \\
       &            & $''$ & $\frac{\rm{mJy}}{\rm{beam}}$ & $\frac{\rm{mJy}}{\rm{beam}}$  \\
\hline
G305.21+0.21 & NH$_3$(4,4) & $0.81\times 0.56$ & 2.5 & 29 \\
G305.21+0.21 & NH$_3$(5,5) & $0.80\times 0.55$ & 2.4 & 31 \\
G305.21+0.21 & cont. & $2.2\times 2.4$ & 0.8 \\
\hline
G316.81-0.06 & NH$_3$(4,4) & $1.83\times 1.25$ & 3.7 & 36 \\
G316.81-0.06 & NH$_3$(5,5) & $2.01\times 1.41$ & 2.6 & 33 \\
G316.81-0.06 & cont. & $1.64\times 1.01$ & 3.7 \\
\hline
G323.74-0.26 & NH$_3$(4,4) & $0.86\times 0.54$ & 2.0 & 19 \\
G323.74-0.26 & NH$_3$(5,5) & $0.81\times 0.51$ & 2.1 & 26 \\
G323.74-0.26 & cont. & $2.61\times 1.56$ & 0.2 \\
\hline
G327.3-0.60 & NH$_3$(4,4) & $0.90\times 0.53$ & 2.6 & 49 \\
G327.3-0.60$^b$ & NH$_3$(4,4) & $2.08\times 1.11$ & 2.1 & 123 \\
G327.3-0.60 & NH$_3$(5,5) & $0.89\times 0.52$ & 2.3 & 54 \\
G327.3-0.60$^b$ & NH$_3$(5,5) & $2.01\times 1.08$ & 2.6 & 140 \\
G327.3-0.60 & cont & $2.64\times 1.47$ & 1.8 &  \\
\hline
G328.81+0.63 & NH$_3$(4,4) & $0.68\times 0.44$ & 2.0 & $-82^e$ \\
G328.81+0.63 & NH$_3$(5,5) & $0.67\times 0.43$ & 1.8 & $-51^e$ \\
G328.81+0.63 & cont. & $0.61\times 0.41$ & 2.8  \\
\hline
G331.28-0.18 & NH$_3$(4,4) & $0.91\times 0.58$ & 1.5 & 22\\
G331.28-0.18 & NH$_3$(5,5) & $0.90\times 0.57$ & 1.5 & 16\\
G331.28-0.18 & cont. & $0.63\times 0.4$ & 0.3  \\
\hline
G336.02-0.83$^b$ & NH$_3$(4,4) & $2.20\times 1.28$ & 2.5 & 21 \\
G336.02-0.83$^b$ & NH$_3$(5,5) & $2.17\times 1.23$ & 2.6 & 20 \\
G336.02-0.83$^b$ & cont. & $2.72\times 1.32$ & 0.35  \\
\hline
G345.00-0.22 & NH$_3$(4,4) & $1.14\times 0.55$ & 1.9 & 43 \\
G345.00-0.22 & NH$_3$(5,5) & $1.12\times 0.54$ & 2.0 & 37 \\
G345.00-0.22 & cont. & $0.76\times 0.34$ & 1.6 \\
\hline
G351.77-0.54 & NH$_3$(4,4) & $1.11\times 0.57$ & 4.0 & 78 \\
G351.77-0.54$^b$ & NH$_3$(4,4) & $2.66 \times 1.21$ & 2.8 & 120 \\
G351.77-0.54 & NH$_3$(5,5) & $1.09\times 0.56$ & 4.1 & 63 \\
G351.77-0.54$^b$ & NH$_3$(5,5) & $2.59 \times 1.19$ & 2.1 & 120 \\
G351.77-0.54 & cont & $1.11\times 0.63$ & 1.1 &  \\
G351.77-0.54$^b$ & cont & $3.08\times 1.55$ & 1.8 &  \\
\hline
G0.55-0.85$^c$ &  NH$_3$(4,4) & $2.10\times 0.75$ & 2.6 & 44 \\
G0.55-0.85$^c$ &  NH$_3$(5,5) & $1.81\times 0.54$ & 2.5 & 32 \\
G0.55-0.85$^c$ &  cont & $1.46\times 0.54$ & 1.3 & 23 \\
\hline
G19.47-0.17$^c$ &  NH$_3$(4,4) & $4.51\times 0.85$ & 3.1 & 17$^d$ \\ 
G19.47-0.17$^c$ &  NH$_3$(5,5) & $6.48\times 1.18$ & 4.8 & 17$^d$ \\ 
G19.47-0.17$^c$ &  cont & $11.92\times 1.05$ & 0.35 & 6 \\ 
\hline
I18151-1208 & NH$_3$(4,4) & $2.60\times 0.58$ & 2.9 & \\ 
I18151-1208 & NH$_3$(5,5) & $2.53\times 0.58$ & 4.1 & \\ 
I18151-1208 & cont & $2.60\times 0.58$ & 0.24 & 1.22 \\ 
\hline
\end{tabular}
{\footnotesize 
\noindent $^a$ For the line rms we used channel separations of 0.8\,km\,s$^{-1}$.\\
$^b$ Only antennas 1 to 5 were used for these lower-resolution images.\\
$^c$ Only limited shorter baseline ranges were used to produce these images.\\
$^d$ Peak flux of integrated image from 17 to 23\,km\,s$^{-1}$.\\
$^e$ Negative because in absorption.
}
\label{tablesample2}
\end{table}

\section{Sample}
\label{sample}

The source sample was largely identified by previous lower resolution
NH$_3$ studies of 41 sources using the ATCA \citep{longmore2007} and
60 sources using Mopra (Walsh et al.~priv. comm.). All sources were
selected based on strong NH$_3$ (4,4) and (5,5) emission, they show
outflow signatures and are prominent in other dense core tracers
(e.g., \citealt{purcell2006,purcell2009}).  IRAS\,18151-1208 does not
follow these identification criteria but was selected because of
recent disk-like structures observed in the 1.3\,mm continuum emission
(Fallscheer et al.~in prep.).  The outflow orientations, which have to
be perpendicular to the expected disks, are known for a majority of
the sources.  The studied sources comprise the best
high-mass-disk-candidate sample for the southern hemisphere to date.
The sample size (12 sources) was chosen because it doubles the number
of existing disk candidates as listed by \citet{cesaroni2006}, which
were observed heterogenously by different groups, with different
tracers and different selection criteria. In contrast to that, these
new data provide a homogeneous dataset which is easier to interpret.
Investigating such a large sample is the only way to characterize
high-mass disk properties in a general way, important for a
comprehensive understanding of high-mass star formation.

\section{Results}

We will first present the observational results for each source
individually and then put them into a general context in section
\ref{general}.

\subsection{Results for individual sources}

\subsubsection{G305.21+0.21 (IRAS\,13079-6218)} 

\begin{figure*}
\centering
\includegraphics[width=0.4\textwidth,angle=-90]{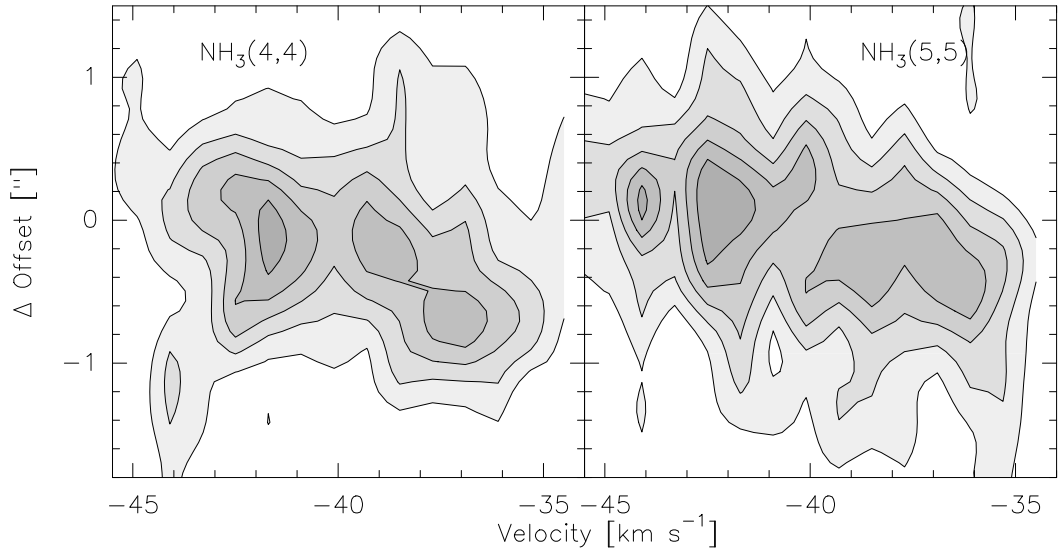}
\caption{G305.21+0.21 position velocity diagrams of the main hyperfine
  components of NH$_3$(4,4) (left) and NH$_3$(5,5) (right). The
  diagrams are centered at Offsets $(0''/0'')$ with a position angle
  of -45 degrees from north (northwest-southeast direction).}
\label{g305_pv}
\end{figure*}

This region exhibits linear maser features with an approximate
northeast-southwest direction \citep{norris1993} similar to the H$_2$
emission features by \citet{debuizer2003}. We clearly detected the
NH$_3$(4,4) and (5,5) lines including their hyperfine structure
components, and as shown in Figure \ref{g305_mom1}, the corresponding
intensity-weighted velocity maps (1st moment maps) exhibit a velocity
gradient in the northwest-southeast direction, approximately perpendicular
to the maser and H$_2$ signatures. The extent of the velocity
structure is approximately $2''$ corresponding at the adopted distance
of 3.5\,kpc to a size of $\sim$7000\,AU. Are these NH$_3$
signatures due to rotation from an infalling/rotating envelope and/or
an embedded high-mass accretion disk? The pv-diagrams of the NH$_3$(4,4)
and (5,5) lines in Figure \ref{g305_pv} also clearly show the velocity
gradient, however, a profile similar to a Keplerian disk is hardly
discernable. Hence, it is more likely that the NH$_3$ structure
corresponds to a large rotating and infalling envelope that may feed a
real Keplerian accretion disk at its center.

Figure \ref{g305_mom1} also presents the intensity-weighted line
width maps (2nd moments). The line width distribution of the (4,4)
transition exhibits several positions of increased line width. This is
likely due to a clumpy sub-structure of the core and the very high
optical depths of the line (see also Fig.~\ref{spectra1} for example
spectra). The line width distribution of the (5,5) transition is
simpler with a clear line width increase toward the center. This
indicates that the higher excited line
($E_{\rm{lower}}(\rm{NH}_3(5,5))=295$\,K) probes deeper into the core
center better allocating the position of the central protostar. Such a
central line width increase is also necessary in the picture of a
rotating, infalling envelope with a central disks.

\begin{figure}
\includegraphics[width=0.3\textwidth,angle=-90]{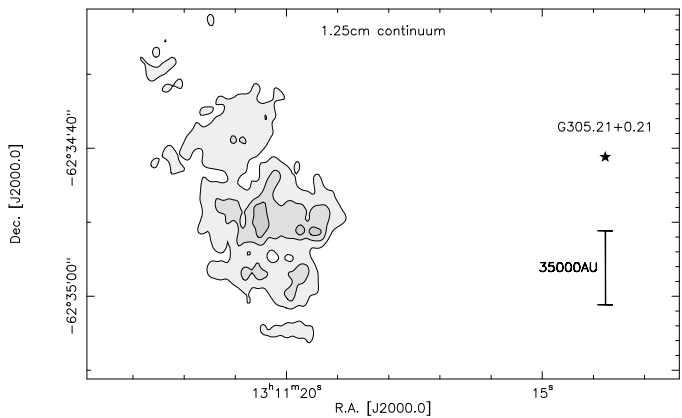}
\caption{1.25\,cm continuum emission in the field of G305.21+0.21. The
  contour levels are in $3\sigma$ steps (Table \ref{tablesample2}).
  The star marks the position of our primary NH$_3$ target within the
  field, and a scale-bar is shown at the bottom-right.}
\label{g305_cont}
\end{figure}

We also detect an H{\sc ii} region (Fig.~\ref{g305_cont}), however, that
is located $\sim 50''$ to the east and unlikely to be associated with
our region of interest.

\subsubsection{G316.81-0.06  (IRAS\,14416-5937)} 

This region exhibits Class II CH$_3$OH masers aligned in a north-south
direction \citep{walsh1998} as well as elongated Spitzer 4.5\,$\mu$m
``green fuzzy'', extended Ks-band emission (likely from shocked H$_2$)
and 7\,mm continuum emission with an orientation approximately NNW-SSE
\citep{longmore2009,longmore2009b}. While the Spitzer 4.5\,$\mu$m
emission is usually attributed to shocked H$_2$ from an underlying
outflow/jet system (e.g., \citealt{noriega2004}), 7\,mm continuum
emission could in general be attributed to jet or disk emission (e.g.,
\citealt{zapata2006a,araya2007}).

For this region, we do detect the NH$_3$(4,4) and (5,5) lines less
strongly than, e.g., toward the previously discussed region
G305.21+0.21, and we had to image them with a lower spatial resolution
(see Table \ref{tablesample2}). However, the structure is clearly
elongated approximately in a north-south direction
(Fig.~\ref{g316_mom1}). While the Spitzer 4.5\,$\mu$m emission
co-alignment with the NH$_3$ features would suggest an association
with the outflow, this appears less likely for spectral lines with
high excitation temperatures (exceeding 200\,K, see \S\ref{intro}). It
appears that for G316.81-0.06 our NH$_3$ data are inconclusive whether
they can be associated with a rotating structure or not. Similarly,
with the given low signal-to-noise ratio the 2nd moment line width
distributions are also inconclusive with regard to the location of the
line width maximum.

We do also detect extended 1.2\,cm continuum emission from an H{\sc
  ii} region, however, that is located approximately $25''$ to the
west and hence not directly associated with our target source
(Fig.~\ref{g316_cont}).

\subsubsection{G323.74-0.26 (IRAS\,15278-5620)} 

The Class II CH$_3$OH maser features do not follow a clear
trend, but most emission peaks appear to be aligned mainly in an
east-west direction \citep{norris1993,walsh1998}. Furthermore,
\citet{walsh2002} identify a H$_2$ outflow structure in approximately
east-west direction.  Spitzer exhibits green fuzzy emission in the
4.5\,$\mu$m band as well, but it is difficult to associate an obvious
outflow direction with that.

Our high-excitation NH$_3$ observations also show no obvious trends,
and the emission features are even slightly different in the (4,4) and
(5,5) line (Fig.~\ref{g323_mom1}). While the (4,4) emission is
elongated in approximately southeast-northwest direction, it does not
show an obvious velocity gradient. In contrast to that, the (5,5)
emission is more compact but shows no clear velocity structure either.
Since the signatures do not coincide in both lines and the outflow
identification is not unambiguous either, we refrain from further
interpretation.

The 2nd moment line width distributions exhibit the line width maximum
close to the main group of Class II CH$_3$OH maser features, hence
indicating that the center of gas infall and star formation activity,
and hence the location of the main protostar is likely close to that.

The 1.25\,cm continuum shows a tentative $5\sigma
=1$\,mJy\,beam$^{-1}$ emission feature $+11.1''/-7.5''$ offset from
the phase center.

\subsubsection{G327.3-0.60} 
\label{g327}

This source was part of the Class II CH$_3$OH maser sample by
\citet{caswell1995}. It is one of the strongest NH$_3$(4,4) and (5,5)
emiting sources in our sample (Table \ref{tablesample2}). With this
strong emission it is also easy to detect and map the hyperfine
structure components. Figure \ref{g327_mom1_noant6} shows the 1st
moment maps of the main hyperfine components of the NH$_3$(4,4) and
(5,5) lines excluding antenna 6 to better show the large-scale
velocity gradient. We clearly identify a velocity gradient in
north-west south-east direction centered around the CH$_3$OH maser
position. 

\begin{figure*}
\centering
\includegraphics[width=0.7\textwidth,angle=-90]{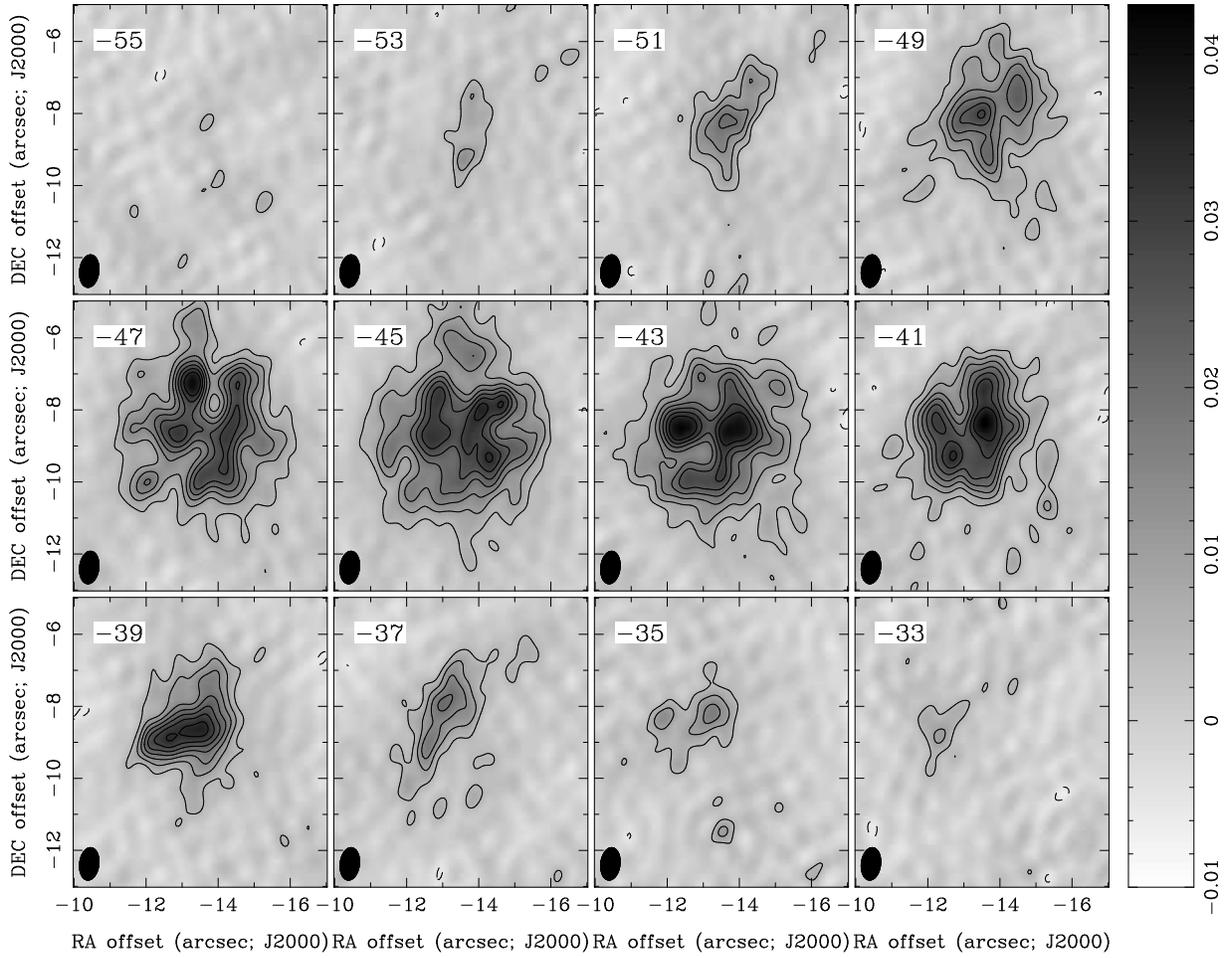}
\caption{Channel map of the main hyperfine component of NH$_3$(4,4)
  with a spectral resolution of 2\,km\,s$^{-1}$ in G327.3-0.60. The
  contour levels (positive full lines, negative dashed lines) are in
  $3\sigma$ steps with a $1\sigma$ value of 1.6\,mJy\,beam$^{-1}$. The
  synthesized beams are shown at the bottom-left of each panel, and
  the 0/0 position is given in Table \ref{tablesample}.}
\label{g327_44_channel}
\end{figure*}

\begin{figure}
\centering
\includegraphics[width=0.4\textwidth,angle=-90]{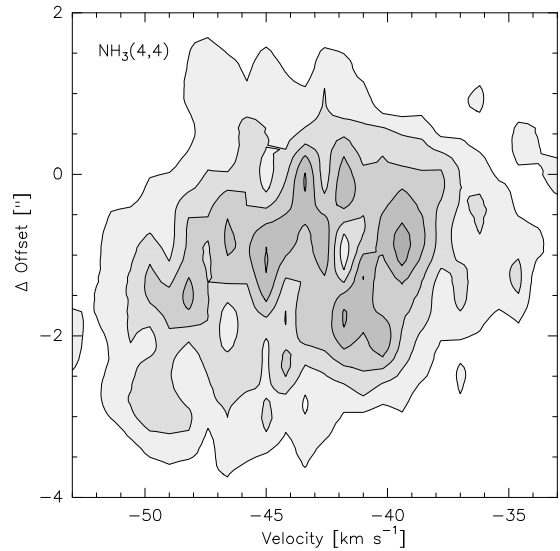}
\caption{G327.3-0.60 position velocity diagram of the main hyperfine
  components of NH$_3$(4,4) (left). The diagram is centered at offset
  $(-12.5''/-9.1'')$ with a position angle of 140 degrees from north
  (northwest-southeast direction).}
\label{g327_pv}
\end{figure}

Including now antenna 6 in the imaging process to also study
smaller-scale sub-structure, Figure \ref{g327_mom1} presents the 1st
moment maps of the main hyperfine components of the NH$_3$(4,4) and
(5,5) lines as well as the 1st moment map corresponding to the most
blue-shifted hyperfine structure line (offset by 2.45\,MHz from the
main line). While the most blue-shifted hyperfine line exhibits a
relatively clear velocity gradient in approximately
northwest-southeast direction, the picture is less clear for the main
hyperfine structure line of both NH$_3$ transitions. Inspection of the
spectra shows that the hyperfine satellite lines have almost the same
intensities as the central main line which indicates the extremely
high optical depth.  With such high optical depth we only see the
outer layers of the rotating and likely collapsing core. In these
lines we also identify a velocity gradient, however, the highest
velocities are not found at the southeastern edge as for the
blue-shifted NH$_3$(4,4) hyperfine line, but it is shifted about $1''$
inward of the core. This feature is reminiscent to the so-called
bull's-eye structure observed in NH$_3$(3,3) absorption lines by
\citet{sollins2005} toward the very luminous high-mass star-forming
region G10.6-0.4. The bull's-eye structure features the gas with the
highest redshift with respect to the $v_{\rm{lsr}}$, and
\citet{sollins2005} interpret this structure as caused by spherical
infall of the core. Similarly, in our source G327.3-0.60, the most
red-shifted feature at -42\,km\,s$^{-1}$ with respect to the
$v_{\rm{lsr}}$ of about $-46$\,km\,s$^{-1}$ (Table \ref{tablesample}
and Figure \ref{g327_mom1} left and middle panel) exhibits the similar
structure where the surrounding gas in all directions features lower
velocities again. Therefore, we may witness here some more spherical
infall motions in the outer core that is traced by the more optically
thick main hyperfine components as well.  To further emphasize the
complex morphological and kinematic structure of this core, Figure
\ref{g327_44_channel} presents a channel map of the region. We clearly
see several distinct clumps which may all be associated with infall
motions. At the near kinematic distance of $\sim$3.1\,kpc (Table
\ref{tablesample}), the spatial extent of the structure
($\sim$5.5$''$) corresponds to an approximate linear extent of
17000\,AU. A position-velocity cut of the NH$_3$(4,4) main hyperfine
line through the bull's-eye feature with a position angle of 140
degrees east of north (Figure \ref{g327_pv} left panel) also
highlights the complexity of the structure without any Keplerian
disk-like signature.

The picture appears slightly different if one investigates the
more optically thin blue-shifted NH$_3$(4,4) hyperfine line (Figure
\ref{g327_mom1} right panel). There we do not see the bull's-eye
structure, but the 1st moment map features a more consistent
velocity gradient in northwest-southeast direction. Since we do not
know for certain about any outflow structure, assigning this velocity
gradient to a rotating disk-like entity is dubious.
Nevertheless, since these highly excited lines are not expected to
trace outflow motions, it is tempting to interpret the NH$_3$(4,4)
hyperfine line 1st moment map as tentative evidence for rotation in
this core. 

Combining the previous spherical infall signature from the more
optically thick main hyperfine structure lines with the tentative
rotational signature of the more optically thin blue-shifted hyperfine
structure line, these data are consistent with a large-scale infalling
envelope that appears spherical in the outer regions and exhibits
stronger rotation signatures further inside due to the conservation of
angular momentum.

Figure \ref{g327_mom1} also presents the corresponding line width 2nd
moment maps. In all three cases the line width distribution peaks
approximately in the center of the rotating structure, offset from the
previously discussed bull's eye feature, and closer to the position of
the Class II CH$_3$OH maser features. This again indicates that the
likely center of star formation activity, and hence the position of
the central protostar, is actually at the center of the core and not
toward the bull's eye position.

We also detect a 1.25\,cm continuum source within our field, however,
this is again more than $10''$ offset from our target and can
therefore be considered in our context as unrelated.

\subsubsection{G328.81+0.63 (IRAS\,15520-5234)} 

As shown in Figure \ref{g328_mom1}, this region exhibits a small
cluster of 1.25\,cm continuum sources with more than 10 emission peaks
within the inner 21000\,AU. The Class II CH$_3$OH masers are located
at the edge of the UCH{\sc ii} region.  The 1.25\,cm peak flux is
136\,mJy\,beam$^{-1}$. While we do not detect NH$_3$ emission from the
central region, both NH$_3$ transitions show strong absorption
features toward the strongest cm continuum peaks.  Figure
\ref{g328_mom1} shows the 1st moment maps of these absorption
features. While the easternmost peak exhibits the absorption peak
around -39.3\,km\,s$^{-1}$, going further to the east we find peak
velocities around -45.9, -44.8 and -44.0\,km\,s$^{-1}$.  Hence, while
there are clear velocity differences between the sources, there is no
consistent velocity gradient across several sub-peaks. A closer
inspection of the velocity structure of individual sub-peak reveals
that -- similar to sources G327.3-0.60 (Section \ref{g327}) or
G10.6-0.4 \citep{sollins2005} -- toward the strongest eastern cm
continuum source the absorbing gas toward the peak is systematically
red-shifted compared to the edge of the peak position (Figure
\ref{g328_mom1_peaks}). In the above discussed scenario (Section
\ref{g327}), this again indicates approximately spherically infalling
gas toward the H{\sc ii} region.  Follow-up observations of hydrogen
recombination lines would be required to test whether the gas
continues to infall through the H{\sc ii} region, and the source may
hence still be accreting, or whether it is stopped before and the
accretion processes have already terminated (e.g.,
\citealt{keto2002a}).

The line width distributions of the absorption features are shown in
Figure \ref{g328_mom2}, and we clearly see a line width increase
toward the continuum peak positions. There are additional line width
maxima in the Figure between the two main continuum peaks and at the
northern edge of the strongest feature, corresponding to the most
negative velocity features seen in Figure \ref{g328_mom1_peaks}.
However, we refrain from a further interpretation of these edge
effects. 

\begin{figure}
\includegraphics[width=0.45\textwidth,angle=-90]{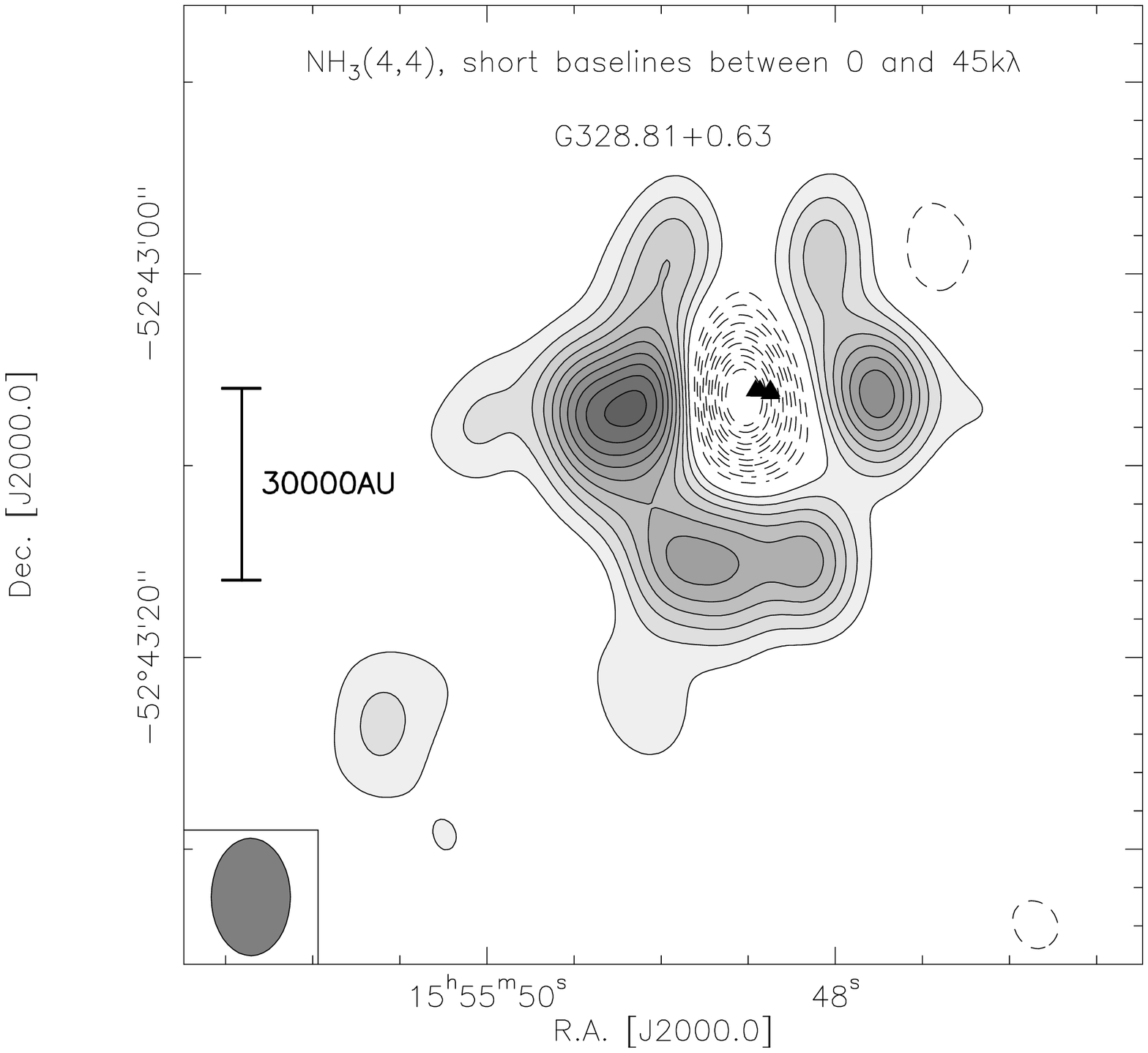}
\includegraphics[width=0.48\textwidth,angle=-90]{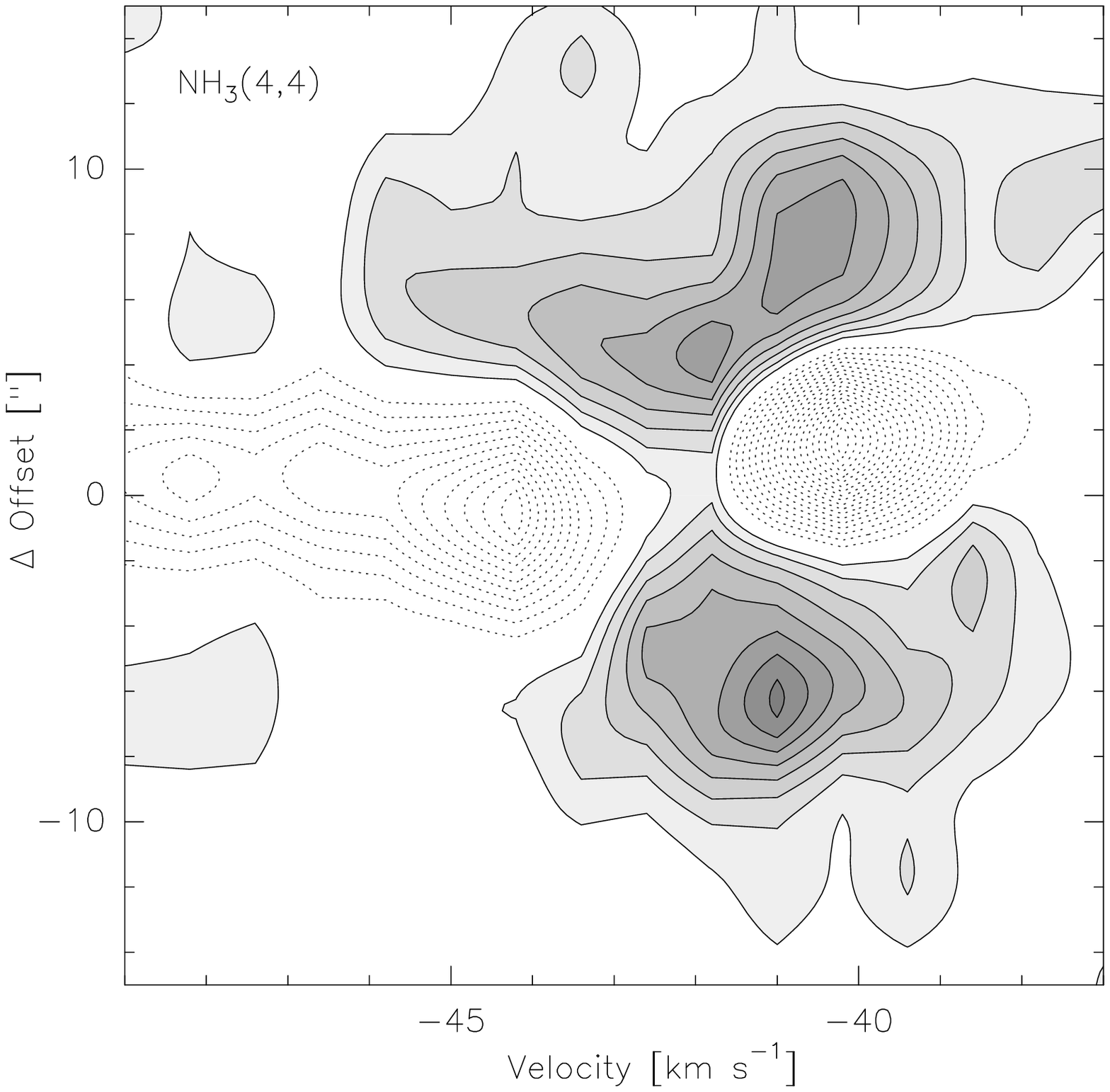}
\caption{G328.81+0.63: {\bf Top:} Integrated NH$_3$(4,4) emission from
  -47.5 to -37.5\,km\,s$^{-1}$. To produce this map, we only used the
  short baselines between 0 and 45\,$k\lambda$. The contouring is done
  in $\pm 3\sigma$ steps of 2.7\,mJy\,beam$^{-1}$. The triangles mark
  the Class II CH$_3$OH maser positions, the synthesized beam and the
  scale-bar are plotted at the left. {\bf Bottom:} The corresponding
  position-velocity diagram through the center in east-west direction.
  The contouring is done in $3\sigma$ levels of 6.6\,mJy\,beam,
  measured in an 0.8\,km\,s$^{_1}$ channel. Full lines show emission,
  dotted lines absorption.}
\label{g328_nh3_large}
\end{figure}

It is interesting to note that although we do not detect NH$_3$ in
emission from the region covered by the UCH{\sc ii} regions, we do
indeed detect strong NH$_3$ emission in both transitions on larger
scales around the UCH{\sc ii} region. Figure \ref{g328_nh3_large}
shows the integrated NH$_3$(4,4) emission of the region at lower
spatial resolution (we only used the baselines between 0 and
45\,$k\lambda$with a synthesized beam of $6.1''\times 4.1''$), and we
clearly identify a ring-like structure with a diameter exceeding
$20''$. Since these are interferometer observations filtering out the
largest spatial scales, it is likely that the real emission is even
more extended. While extended NH$_3$ emission around forming high-mass
stars has previously been observed, for example toward NGC\,6334\,I(N)
it was only seen in the low-energy (1,1) and (2,2) transitions,
whereas the high-energy transitions up to the (6,6) line were observed
only toward the central sources \citep{beuther2005e,beuther2007b}.
Therefore, it is surprising that we witness large-scale hot NH$_3$
emission around G328.81+0.63 from lines with excitation temperatures
as high as 295\,K (see section \ref{intro}). This implies that the
central formed/forming high-mass stars have heated up significant
amounts of gas out to distances exceeding 30000\,AU from the center to
temperatures in excess of 100\,K without yet destroying the
surrounding gas envelope. This may be interpreted as support for the
proposal by \citet{krumholz2006b} that radiative feedback from the
central protostar is able to heat up the surrounding envelope strong
enough that further thermal fragmentation will be largely suppressed.

Figure \ref{g328_nh3_large} also presents a position-velocity cut
through the center of the large-scale map in east-west direction.
While the main emission is approximately between -40 and
-42\,km\,s$^{-1}$, one absorption peak is around -40\,km\,s$^{-1}$
while the rest of the absorption is more blue-shifted $\leq
-44$\,km\,s$^{-1}$. Correlating these features with the
higher-resolution absorption maps in Figure \ref{g328_mom1}, the
-40\,km\,s$^{-1}$ component belongs to the eastern absorption peak,
the features around -44 to -45\,km\,s$^{-1}$ to the relatively
extended east-west continuum ridge and the absorption feature at $\sim
-46$\,km\,s$^{-1}$ to the 2nd strongest absorption weak (2nd peak from
west in Fig.~\ref{g328_mom1}). While for the strongest absorption peak
relative motions with respect to the ambient cloud are not
distinguishable, the blue-shifted absorption data for the 2nd
strongest absorption feature are indicative for expansion motion of
the molecular envelope.

\subsubsection{G331.28-0.19 (IRAS\,16076-5134)} 

The 1.25\,cm continuum maps reveal 2 sources, however, the NH$_3$
emission is only associated with the weaker northern source
(Fig.~\ref{g331_mom1}) which is also the Class II CH$_3$OH maser
emitter. The various outflow tracers like H$_2$ emission, Spitzer
``green fuzzies'', SiO emission or 7\,mm continuum emission all
indicate an outflow direction approximately in northeast-southwest
direction (Table \ref{tablesample}). In contrast to that, our
NH$_3$(4,4) and (5,5) observations are more indicative of a velocity
gradient in northwest-southeast direction, approximately perpendicular
to the outflow. It is interesting to note that the Class II CH$_3$OH
maser orientation appears to align closely with the NH$_3$ emission.
One peculiarity of the NH$_3$ data is that the velocity gradients of
the (4,4) and (5,5) lines are in approximate opposite directions.
Since we do not identify clear Keplerian-like signatures, we do not
observe a real accretion disk, however, the orientation of the NH$_3$
velocity gradient is strongly suggestive of rotating material
perpendicular to the outflow axis. This gas may feed an accretion disk
closer to the center of the core.

Figure \ref{g331_mom1} also shows the 2nd moment line width
distribution. While for the (4,4) line the data are less clear, a
broader line width toward the center close to the Class II CH$_3$OH
maser features is observed for the (5,5) line, consistent with a
central location of the protostar and hence the center of active
infall.

\subsubsection{G336.02-0.83 (IRAS\,16313-4840)} 

\begin{figure}
\includegraphics[width=0.43\textwidth,angle=-90]{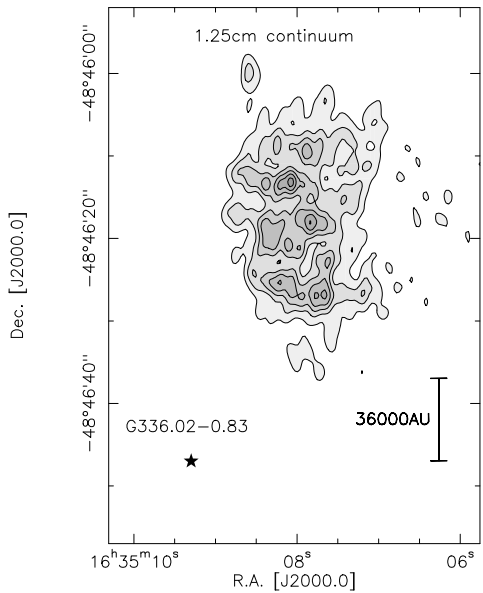}
\caption{1.25\,cm continuum emission in the field of G336.02-0.83. The
  contour levels are at $3\sigma$ intervals (Table
  \ref{tablesample2}).  The star marks the position of our primary
  NH$_3$ target within the field, and a scale-bar adopting the near
  kinematic distance is presented in the bottom-right.}
\label{g336_cont}
\end{figure}

Except for a potential Class II CH$_3$OH maser velocity gradient in
approximately north-south direction \citep{walsh1998}, little else is
known about this region. We do detect weak NH$_3$(4,4) and (5,5)
emission associated with the Class II CH$_3$OH masers, however, only
weakly when excluding the long baselines associated with antenna 6
from the data reduction.  Figure \ref{g336_mom1} shows the 1st moment
maps. Although the spatial resolution does not allow to identify an
obvious velocity gradient, in particular the NH$_3$(5,5) 1st moment
map is indicative of a potential velocity gradient in the north south
direction, parallel to the Class II CH$_3$OH maser features. The 2nd
moment line width distributions also shown in Figure \ref{g336_mom1}
does not exhibit a prominent line width signature, preventing us from
further interpretation. Furthermore, we do detect 1.25\,cm continuum
emission from the region, however, the peak is approximately $30''$
shifted to the north and can be hence considered as unrelated to the
NH$_3$ emission (Fig.~\ref{g336_cont}).

\subsubsection{G345.00-0.22 (IRAS\,17016-4124)} 

Class II CH$_3$OH maser emission is detected toward two positions
approximately $4''$ apart \citep{walsh1998}. While the eastern maser
position is associated with 1.25\,cm continuum emission and
NH$_3$(4,4) and (5,5) absorption, the western maser peak is associated
with NH$_3$(4,4) and (5,5) in emission (Fig.~\ref{g345_mom1}).
Although both maser groups appear to be spatially approximately
aligned with an east-west orientation, they do not show a clear
velocity gradient.  This is at least different for the NH$_3$ emission
toward the western peak position which exhibits a clear velocity
gradient in approximately east-west direction. Toward the eastern
peak, most of the NH$_3$ absorption is blue-shifted with respect to
the $v_{\rm{lsr}}\sim-26.8$ (Table \ref{tablesample}), indicative of
expanding gas. The 2nd moment line width distribution (Figure
\ref{g345_mom2}) toward the western NH$_3$ emission peak shows a line
width increase approximately toward the central Class II CH$_3$OH
emission features. The signatures toward the eastern absorption
features are less conclusive.

Therefore, while the NH$_3$ absorption and 1.25\,cm continuum emission
toward the eastern peak is consistent with an expanding UCH{\sc ii}
region, the NH$_3$ emission data toward the western peak position are
indicative of a rotating structure that is likely still associated
with ongoing high-mass star formation.

\subsubsection{G351.77-0.54 (IRAS\,17233-3606)} 

\begin{figure*}
\centering
\includegraphics[width=0.7\textwidth,angle=-90]{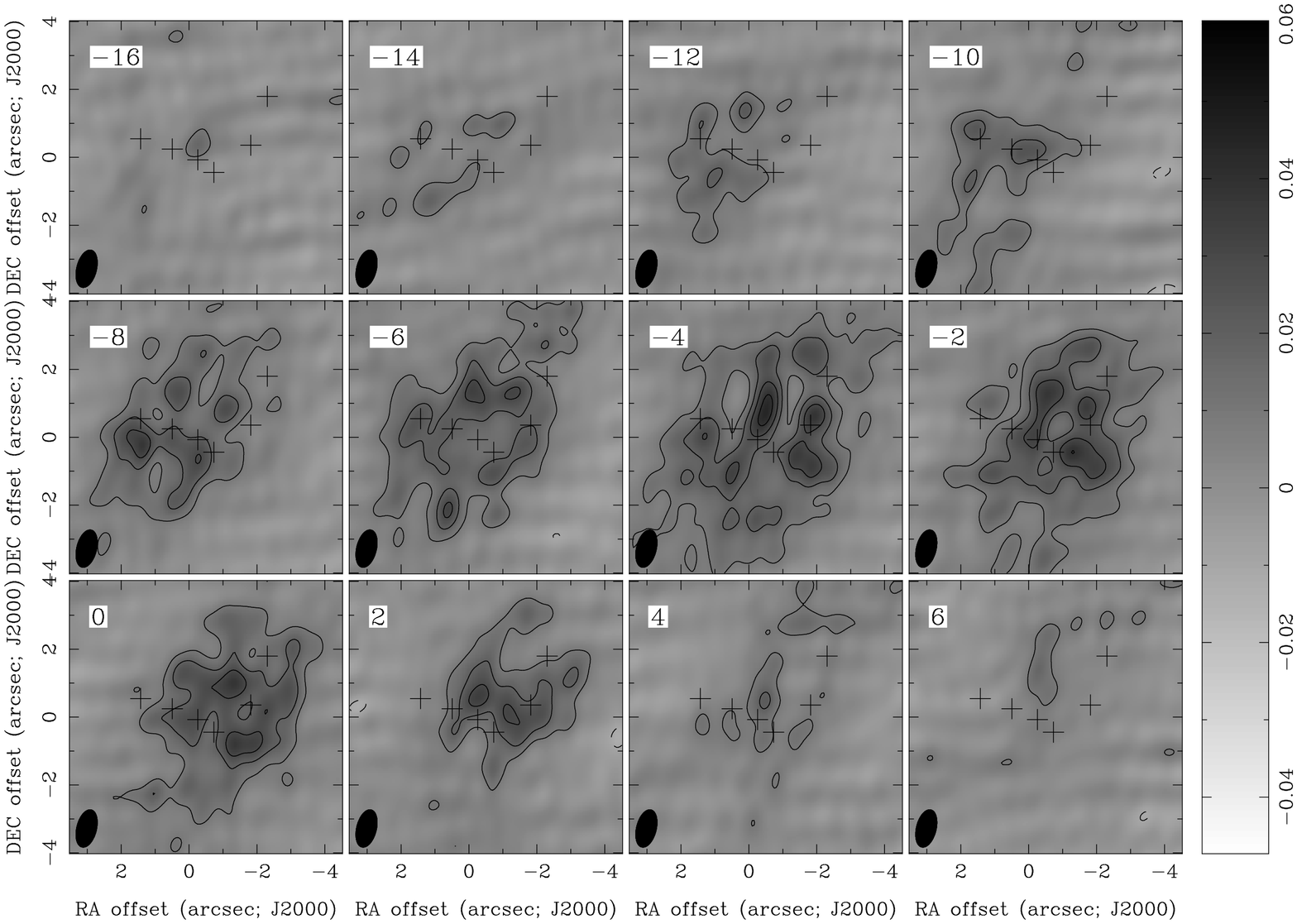}
\caption{Channel map of the main hyperfine component of NH$_3$(4,4)
  with a spectral resolution of 2\,km\,s$^{-1}$ in G351.77-0.54 . The
  contour levels (positive full lines, negative dashed lines) are in
  $3\sigma$ steps with a $1\sigma$ value of 3.3\,mJy\,beam$^{-1}$. The
  crosses mark the cm continuum sources from \citet{zapata2008}, and
  the synthesized beams are shown at the bottom-left of each panel,
  and the 0/0 position is given in Table \ref{tablesample}.}
\label{g351_44_channel}
\end{figure*}

\begin{figure*}
\centering
\includegraphics[width=0.5\textwidth,angle=-90]{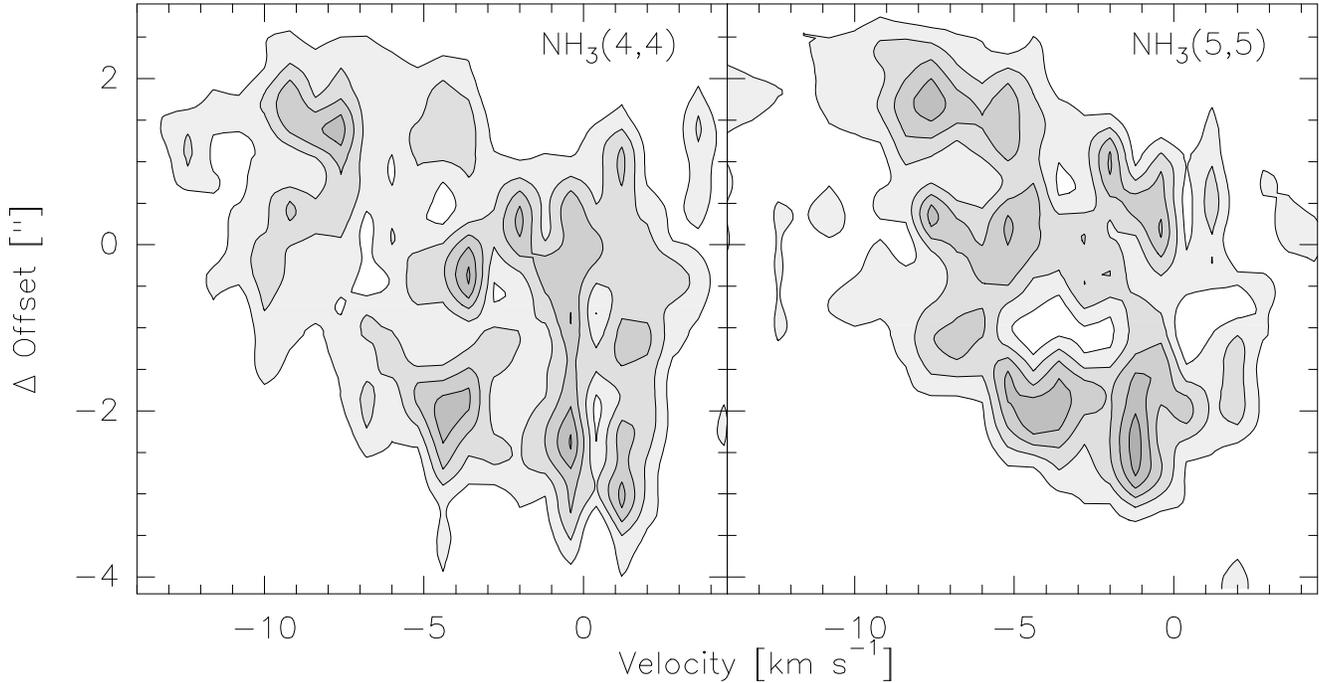}
\caption{G351.77-0.54 position velocity diagrams of the main hyperfine
  components of NH$_3$(4,4) (left) and NH$_3$(5,5) (right). The
  diagrams are centered at Offsets $(0''/0'')$ with a position angle
  of 105 degrees from north (ESE-WNW).}
\label{g351_pv}
\end{figure*}

This region is one of the previously most studied sources in our
sample. It exhibits linear CH$_3$OH maser features approximately in
northeast-southwest direction aligned with a CO outflow of similar
orientation \citep{norris1993,walsh1998,leurini2008}. Furthermore, it
exhibits an OH maser velocity gradient in approximately north-south
direction as well as a H$_2$O maser structure that can be either
interpreted as a ring potentially associated with several recently
identified cm sub-sources (Figs.~\ref{g351_mom1_noant6} and
\ref{g351_mom1}), or as well as a velocity gradient in
northeast-southwest direction \citep{forster1990,fish2005,zapata2008}.
We do detect strong NH$_3$(4,4) and (5,5) emission from the CH$_3$OH
maser position with a clear velocity gradient in ESE-WNW direction,
approximately perpendicular to the outflow and maser orientation. This
signature can be depicted in the lower-resolution image excluding
antenna 6 in the data reduction to highlight the larger-scale rotating
signature (Fig.~\ref{g351_mom1_noant6}), as well as in the
highest-resolution images including antenna 6 to also show
smaller-scale sub-structures (Figs.~\ref{g351_mom1} \&
\ref{g351_44_channel}). This can be interpreted as good evidence of
rotational motion of the core.  Although the channel map in
Fig.~\ref{g351_44_channel} exhibits a clumpy structure of the rotating
gas similar to G327.3-0.60 (section \ref{g327}), the velocity gradient
can also clearly be depicted in the clumpy substructure. While this
may not be significant, it is interesting to note that the VLA cm
continuum sources from \citet{zapata2008} correlate in some spectral
channels with the clumpy molecular sub-structure
(Fig.~\ref{g351_44_channel}). The position-velocity diagram in Figure
\ref{g351_pv} also shows this velocity gradient, however, again the
structure is very clumpy and does not resemble what one would expect
from a Keplerian disk. It more resembles a large-scale rotating
envelope structure that may feed the potential inner accretion disk.
This is consistent with the 2nd moment line width distribution
(Fig.~\ref{g351_mom1}) which is centrally peaked and hence consistent
with increasing rotational velocities toward the center.  This
large-scale rotating envelope has a projected diameter of $\sim 5''$
corresponding at the given distance of $\sim$2.2\,kpc to a an
approximate extent of $\sim$11000\,AU. Since this structure
encompasses all 6 cm continuum sources identified by
\citet{zapata2008}, this rotating envelope may even feed several
smaller independent accretion disks that could be associated with
individual sub-sources. Nevertheless, since the outflow and rotation
signatures do not appear very disturbed by the multiplicity, it is
interesting that the general outflow and perpendicular rotation
structures appear to be dominated by one object.

\begin{figure}
\includegraphics[width=0.48\textwidth,angle=-90]{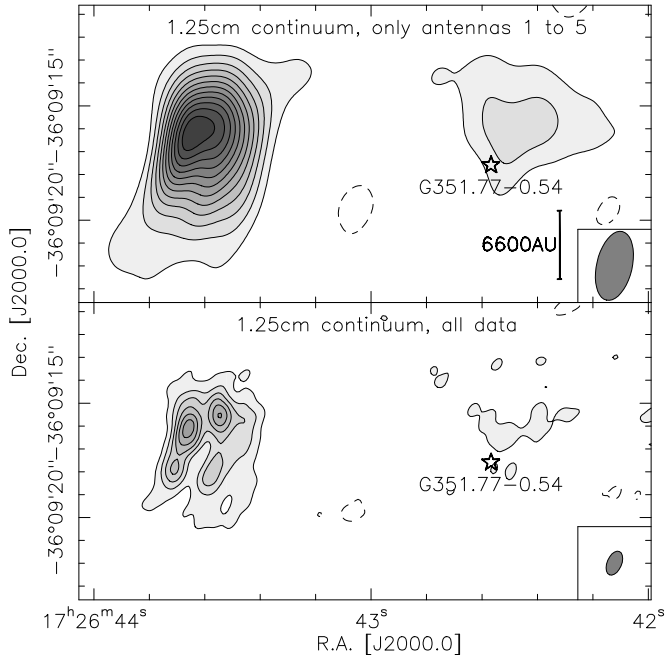}
\caption{1.25\,cm continuum emission in the field of G351.77-0.54. The
  top-panel shows the data excluding the long baselines associated
  with antenna 6, hence with lower spatial resolution (Table
  \ref{tablesample2}). The contour levels are in $3\sigma$ steps
  (Table \ref{tablesample2}, full lines positive, and dashed lines
  negative features).  The star marks the position of our primary
  NH$_3$ target within the field, and a scale-bar is presented in the
  top panel.}
\label{g351_cont}
\end{figure}

We also detect strong 1.25\,cm continuum emission from a nearby
ultracompact H{\sc ii} region separated by about $12''$ to the east
(Fig.~\ref{g351_cont}). At the highest spatial resolution available,
this UCH{\sc ii} region splits up into 4 sub-sources. In addition to
the UCH{\sc ii} region, we detect weak cm continuum emission with a
peak flux of $\sim$14\,mJy\,beam$^{-1}$ toward the NH$_3$ emission
source. This cm structure is not compact but rather extended and hence
neither resembles a jet-like feature nor a hypercompact H{\sc ii}
region.

\subsubsection{G0.55-0.85 (IRAS\,17470-2853)} 

\begin{figure}
\includegraphics[width=0.43\textwidth,angle=-90]{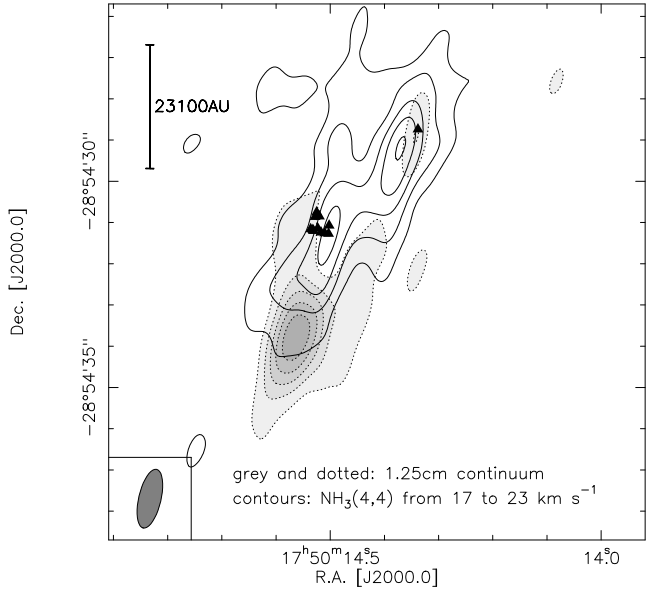}
\caption{The grey-scale with dotted contours shows the 1.25\,cm
  continuum emission in the field of G0.55-0.85. The contour levels
  are at $3\sigma$ intervals (Table \ref{tablesample2}). The full
  contours present the NH$_3$(4,4) emission integrated from 13 to
  21\,km\,s$^{-1}$ with contour levels at $3\sigma$ steps of
  4.8\,mJy\,beam$^{-1}$. The triangles mark the Class II CH$_3$OH
  maser positions by \citet{walsh1998}, the synthesized beam of the
  continuum data is shown at the bottom left, and a scale-bar adopting
  the near kinematic distance is presented in the top-left.}
\label{g055_cont}
\end{figure}

This is again one of the sources with little additional information.
The region hosts two Class II CH$_3$OH maser sites separated by $\sim
3''$ but no obvious velocity gradients are present within them
\citep{walsh1998}. We do detect the NH$_3$(4,4) and (5,5) emission
from both maser positions, and Figure \ref{g055_mom1} presents the
corresponding 1st moment maps of the spectra lines. The two Class II
CH$_3$OH maser feature are clearly connected in the NH$_3$ emission
but neither the (4,4) nor the (5,5) transition exhibits any
conspicuous velocity structure. However, the 2nd moment line width
distribution (Fig.~\ref{g055_mom1} shows a double-peaked structure,
where the two line width peaks are associated with the two Class II
CH$_3$OH maser features and the peaks of integrated NH$_3$ emission
(Fig.~\ref{g055_cont}). This is further evidence for active star
formation activity associated with both emission peaks.

Furthermore, we identify 1.25\,cm continuum emission from a likely
associated UCH{\sc ii} region directly south of the NH$_3$ thermal and
CH$_3$OH maser emission (Fig.~\ref{g055_cont}). In addition to this,
we identify a second 1.25\,cm continuum peak at the $\sim 6\sigma$
level clearly associated with the NH$_3$ peak emission and the
northern CH$_3$OH maser position.

\subsubsection{G19.47+0.17 (IRAS\,18232-1154)} 

\begin{figure} 
\includegraphics[width=0.6\textwidth,angle=-90]{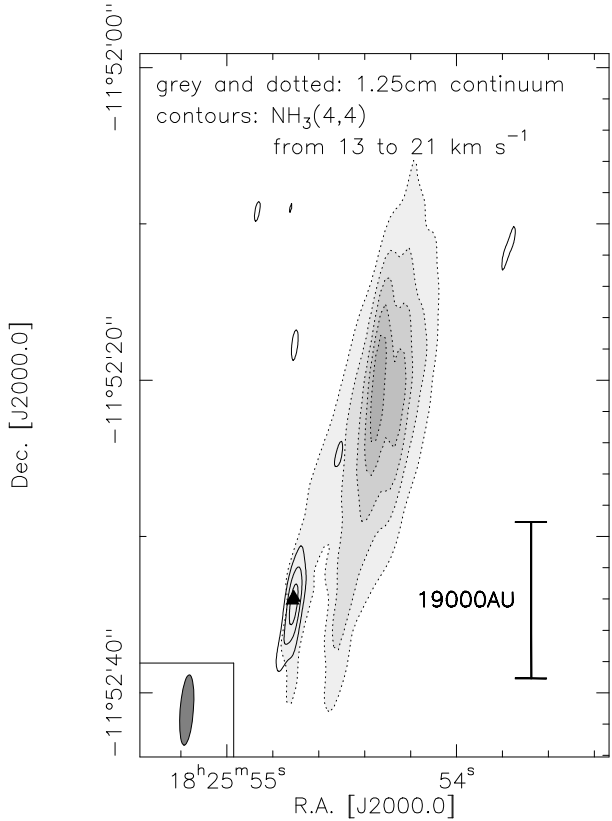}
\caption{The grey-scale with dotted contours shows the 1.25\,cm
  continuum emission in the field of G19.47+0.17. The contour levels
  are at $3\sigma$ intervals (Table \ref{tablesample2}). The full
  contours present the NH$_3$(4,4) emission integrated from 17 to
  23\,km\,s$^{-1}$ with contour levels starting at the $4\sigma$ level
  of 5.6\,mJy\,beam$^{-1}$ and continuing at $3\sigma$ steps. The
  triangles mark the Class II CH$_3$OH maser positions
  \citep{walsh1998}, the synthesized beam of the NH$_3$ data is shown
  at the bottom left, and a scale-bar is presented in the
  bottom-right.}
\label{g19_cont}
\end{figure}

This region exhibits a CO outflow in approximate NNE-SSW direction
(\citealt{longmore2007}, Longmore et al.~in prep.). We do detect both
NH$_3$ lines close to the Class II CH$_3$OH maser position of this
region as well (Table \ref{tablesample}), however, the signal-to-noise
ratio is relatively poor, prohibiting good quality moment maps and
hence deriving reliable velocity gradients. Furthermore, we do detect
extended 1.25\,cm continuum emission toward the north of the NH$_3$
emission peak.  Figure \ref{g19_cont} presents an overlay of the
1.25\,cm continuum emission with the integrated NH$_3$(4,4) emission.
The integrated NH$_3$(4,4) emission map shows several additional
$4\sigma$ features distributed in the vicinity which may indicate more
extended NH$_3$ emission just filtered out by our interferometer
observations (see also \citealt{longmore2007}).

\subsubsection{IRAS\,18151-1208} 

\begin{figure} 
\includegraphics[width=0.6\textwidth,angle=-90]{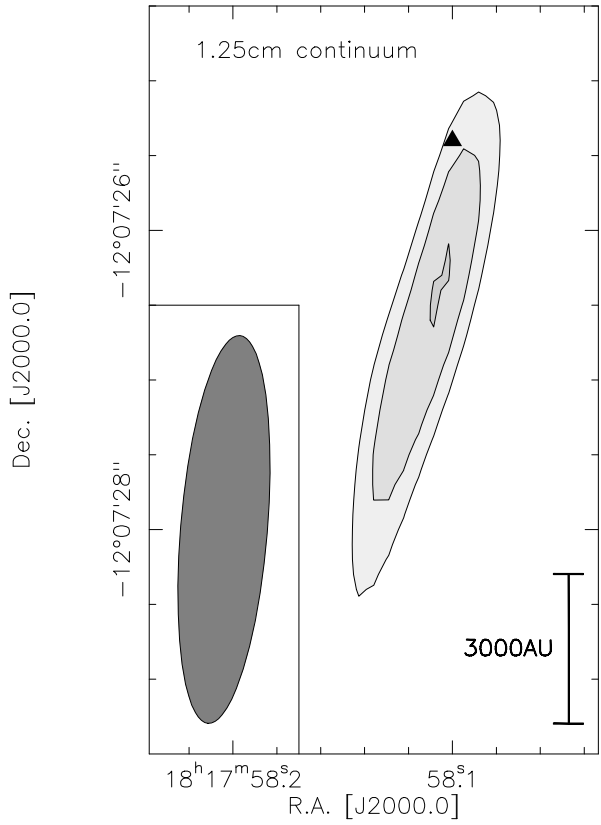}
\caption{1.25\,cm continuum emission in the field of IRAS\,18151-1208.
  The contour levels start at the $3\sigma$ level and continue in
  $1\sigma$ steps (Table \ref{tablesample2}). The triangle marks the
  Class II CH$_3$OH maser position from \citet{beuther2002c}, the
  synthesized beam is shown at the bottom left, and a scale-bar is
  presented in the bottom-right.}
\label{18151_cont}
\end{figure}

As outlined in section \ref{sample}, this region does not satisfy the
selection criteria of the rest of the sample. However, since it is
also a Class II CH$_3$OH maser source and exhibits strong outflow and
disk signatures (\citealt{beuther2002c,beuther2002b,davis2004},
Fallscheer et al.~in prep.), we considered it a good addition for this
observing run.  However, unfortunately it remained undetected in both
NH$_3$ transitions. Nevertheless, we did detect at the $5\sigma$
confidence level 1.25\,cm continuum emission associated with the
CH$_3$OH Class II maser peak (Fig.\,\ref{18151_cont} and Table
\ref{tablesample2}).

\subsection{Spectral fitting and temperature determination}

\begin{figure*} 
\includegraphics[width=0.275\textwidth,angle=-90]{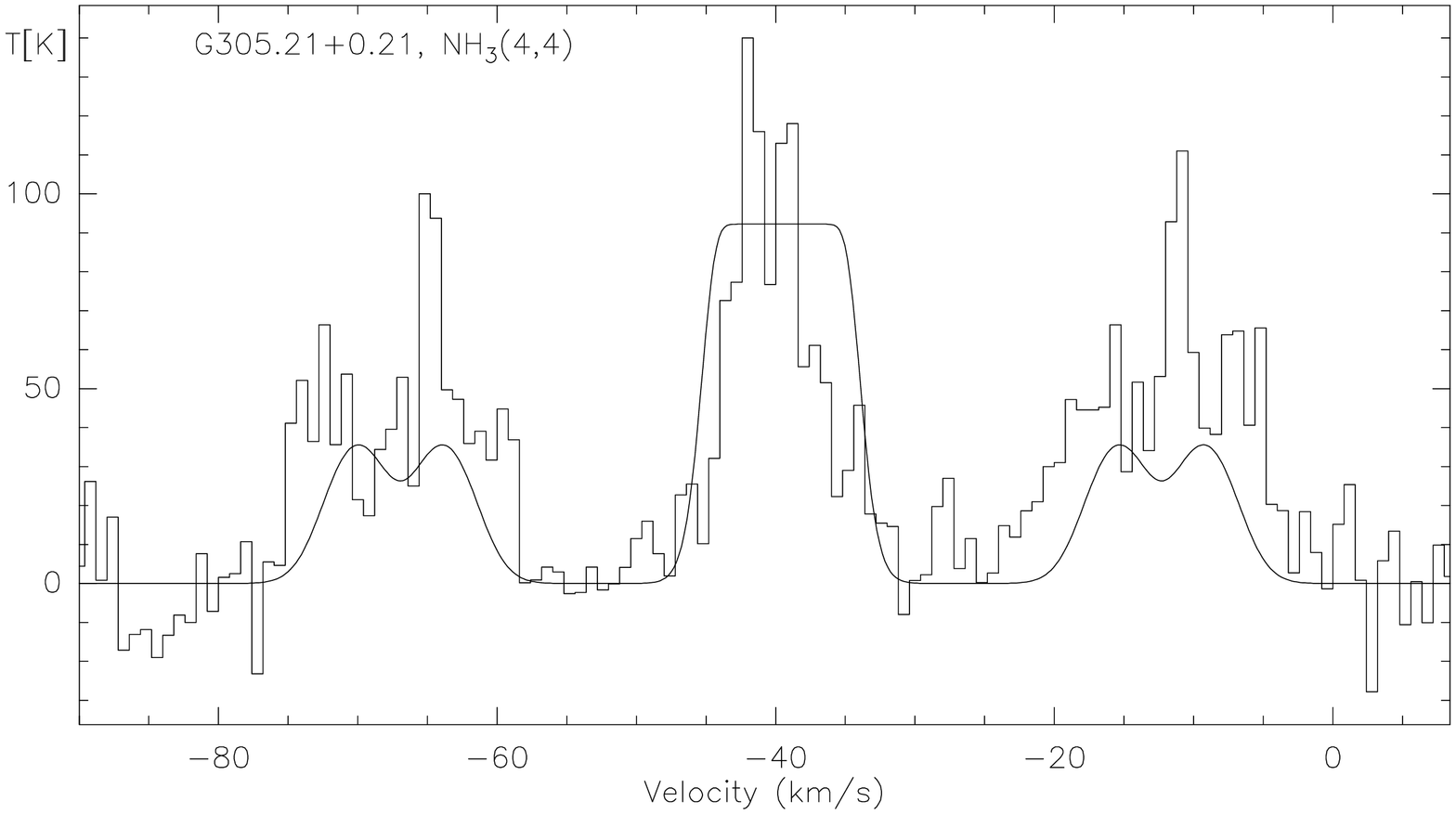}
\includegraphics[width=0.275\textwidth,angle=-90]{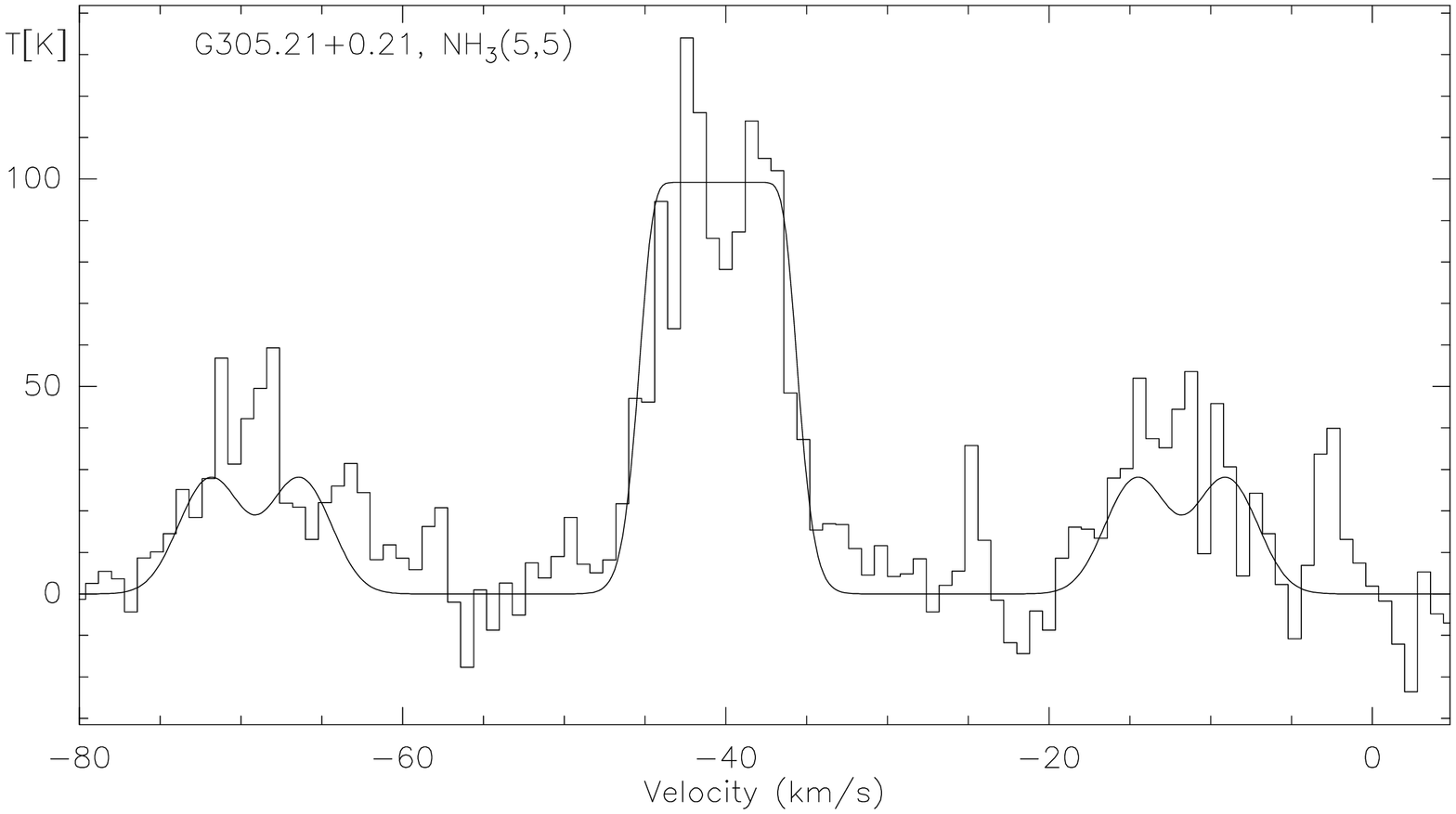}\\
\includegraphics[width=0.275\textwidth,angle=-90]{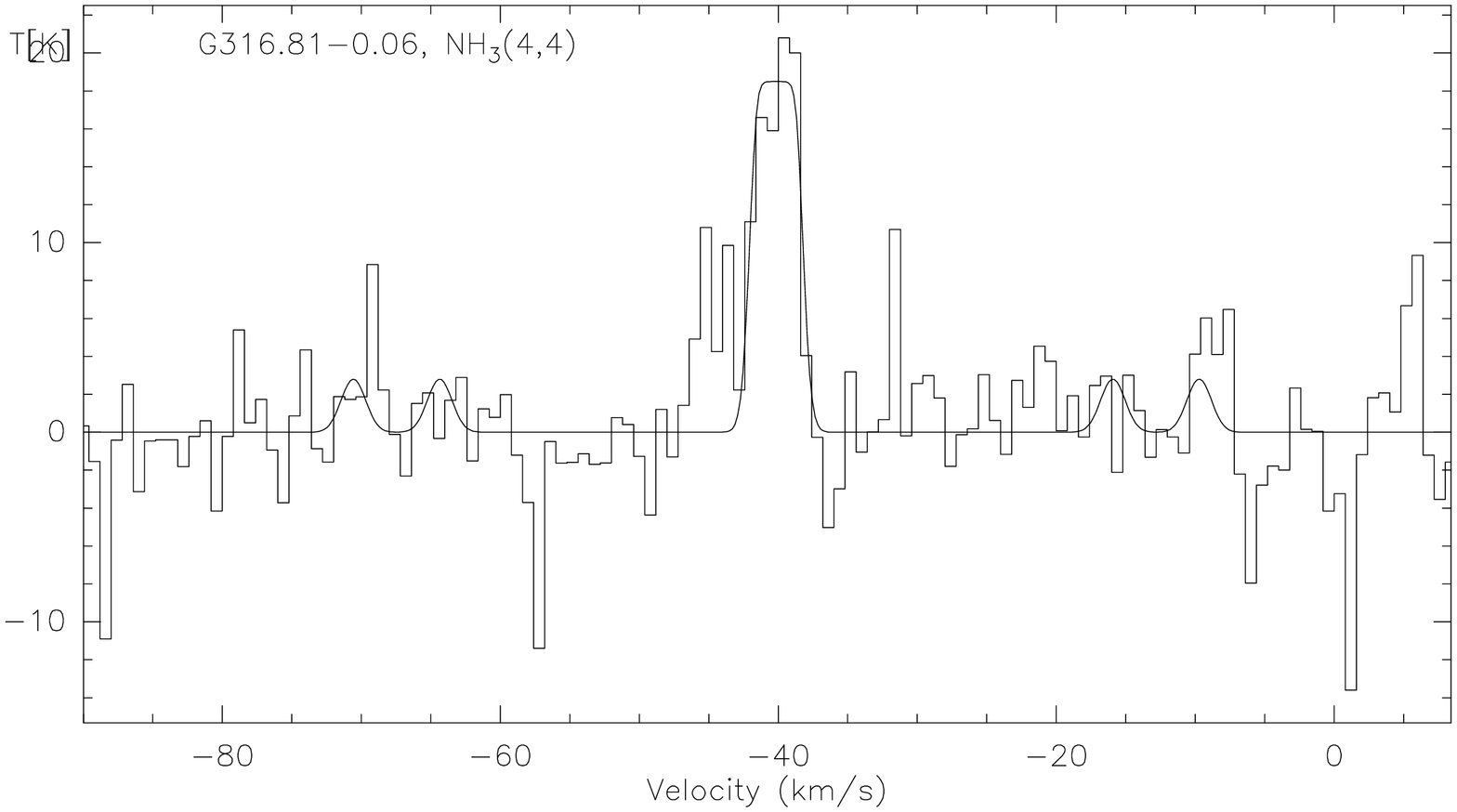}
\includegraphics[width=0.275\textwidth,angle=-90]{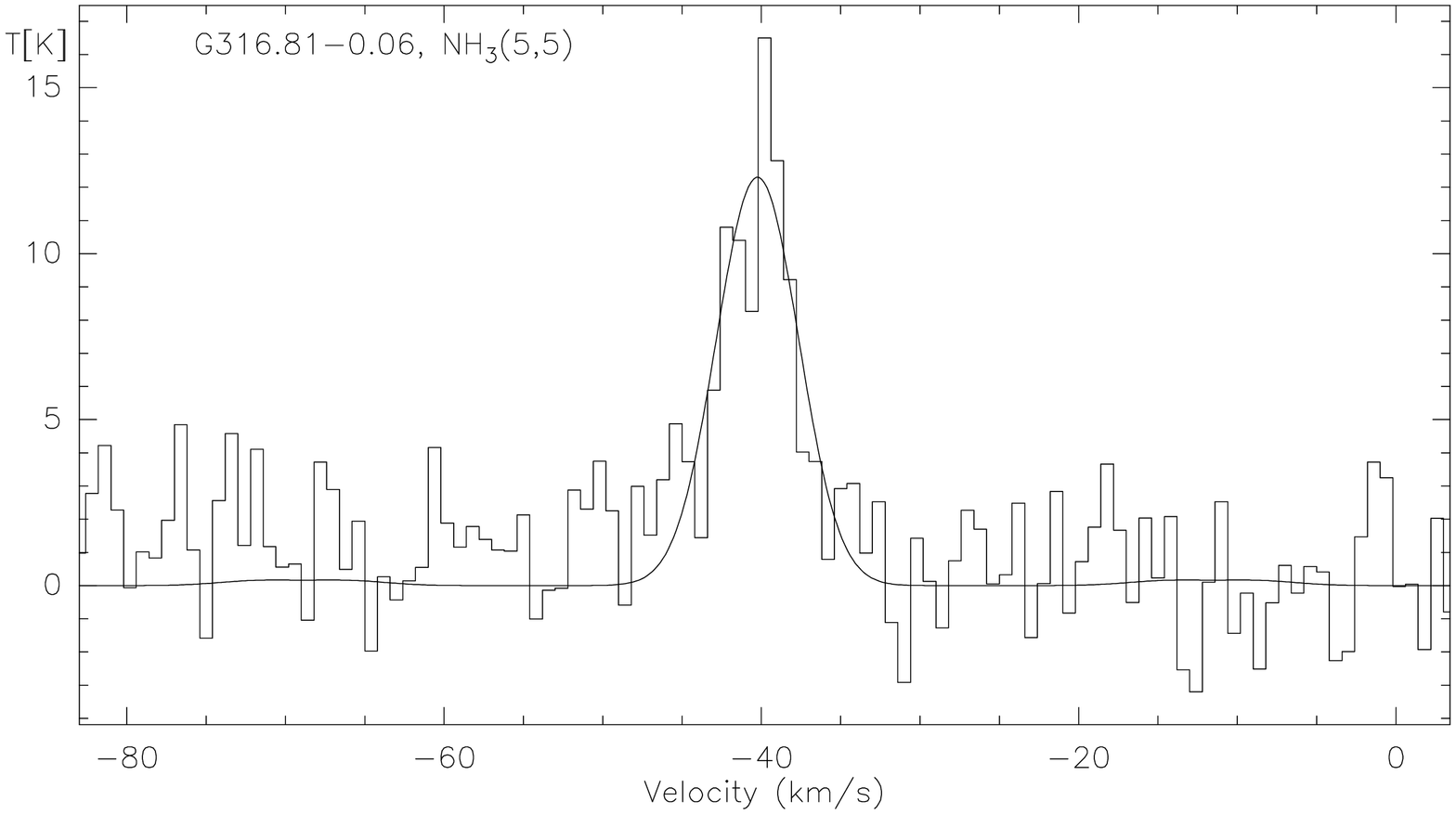}\\
\includegraphics[width=0.275\textwidth,angle=-90]{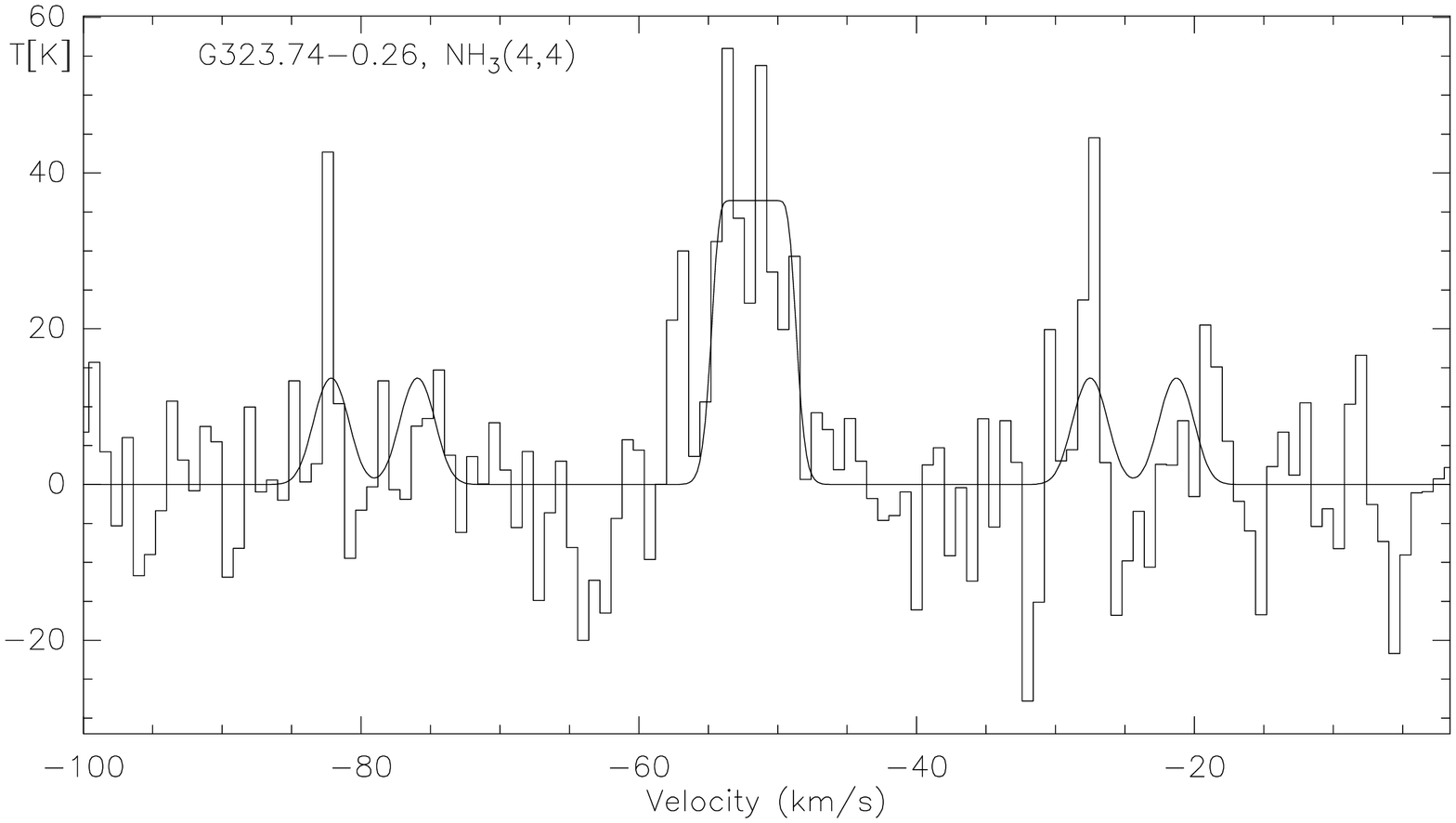}
\includegraphics[width=0.275\textwidth,angle=-90]{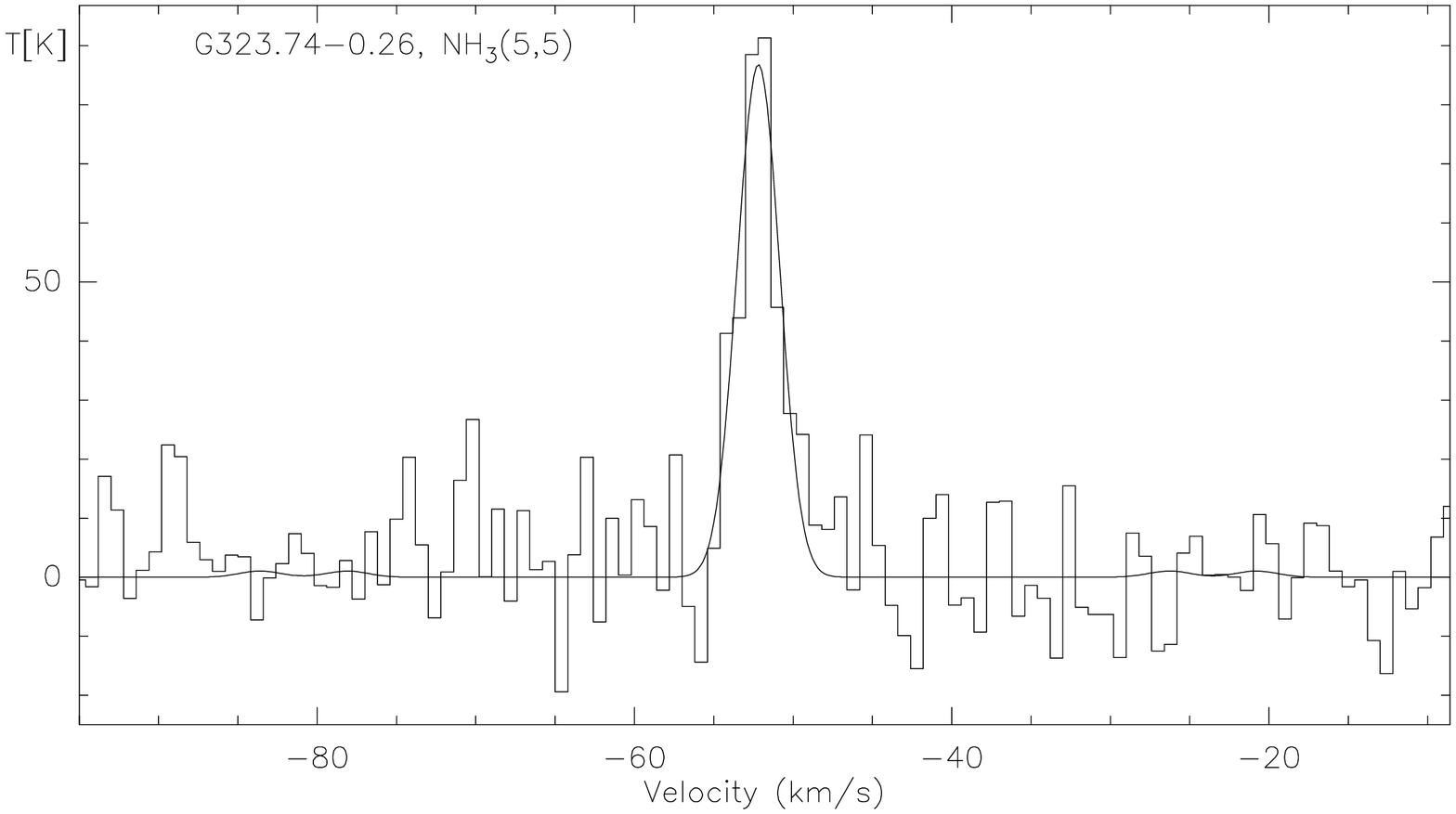}\\
\includegraphics[width=0.275\textwidth,angle=-90]{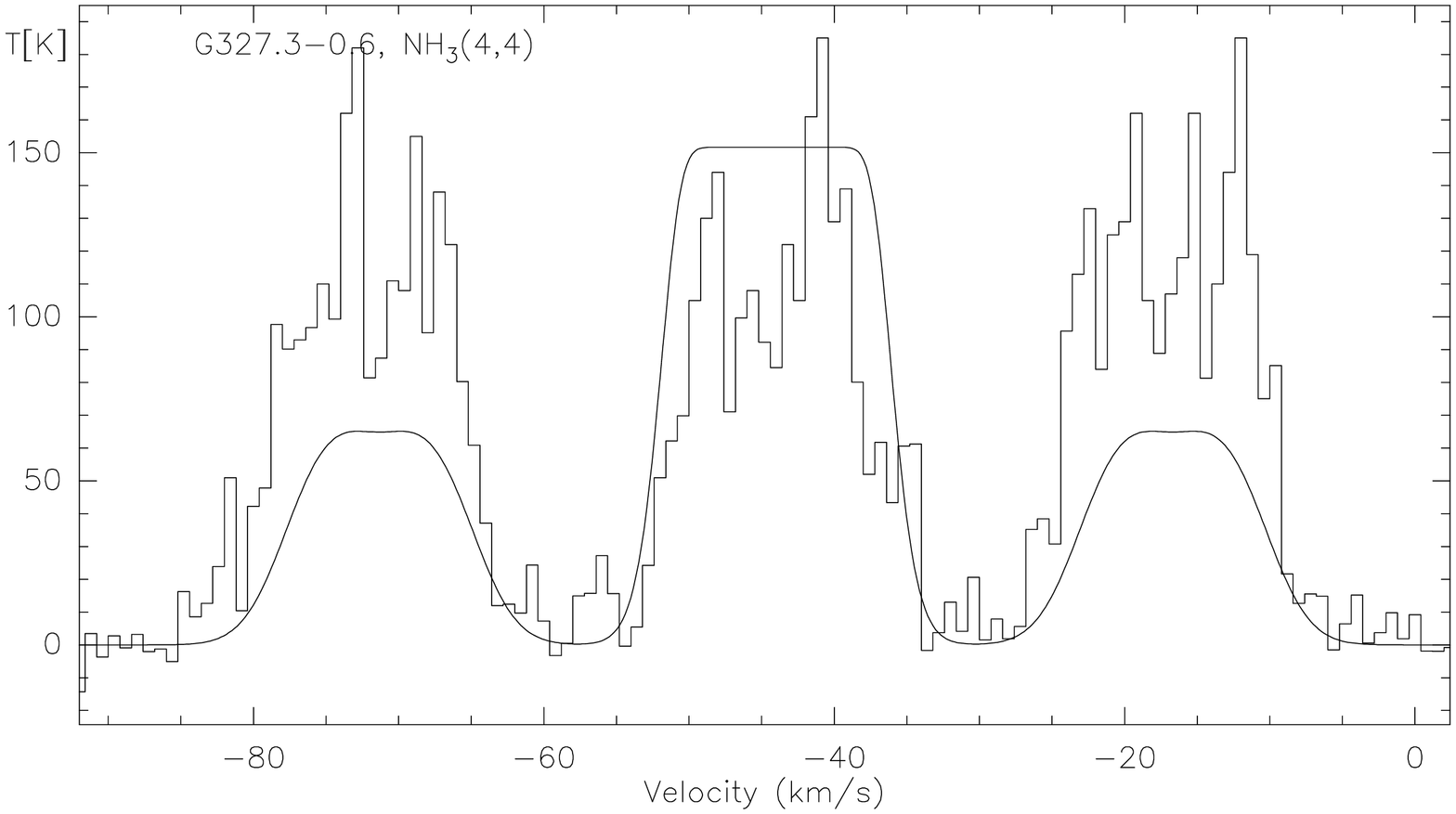}
\includegraphics[width=0.275\textwidth,angle=-90]{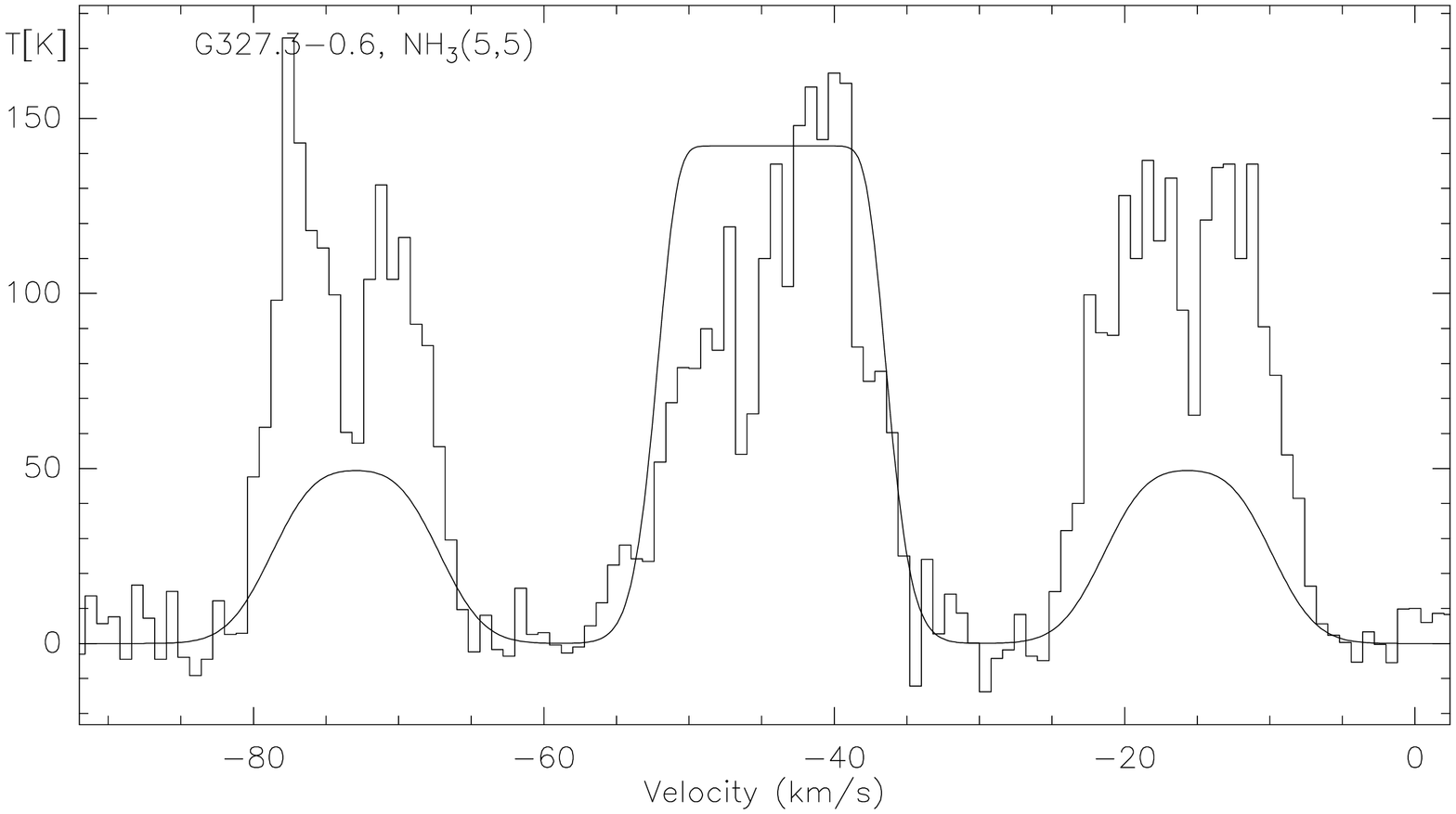}
\caption{NH$_3$(4,4) and NH$_3$(5,5) spectra (left and right column)
  extracted toward the peak positions of the sources labeled in each
  panel. The histogram presents the data whereas the full lines show
  attempts to fit the whole hyperfine structure.  Due to the very high
  optical depth, even this hyperfine structure fitting does not work
  well.}
\label{spectra1}
\end{figure*}

\begin{figure*} 
\includegraphics[width=0.275\textwidth,angle=-90]{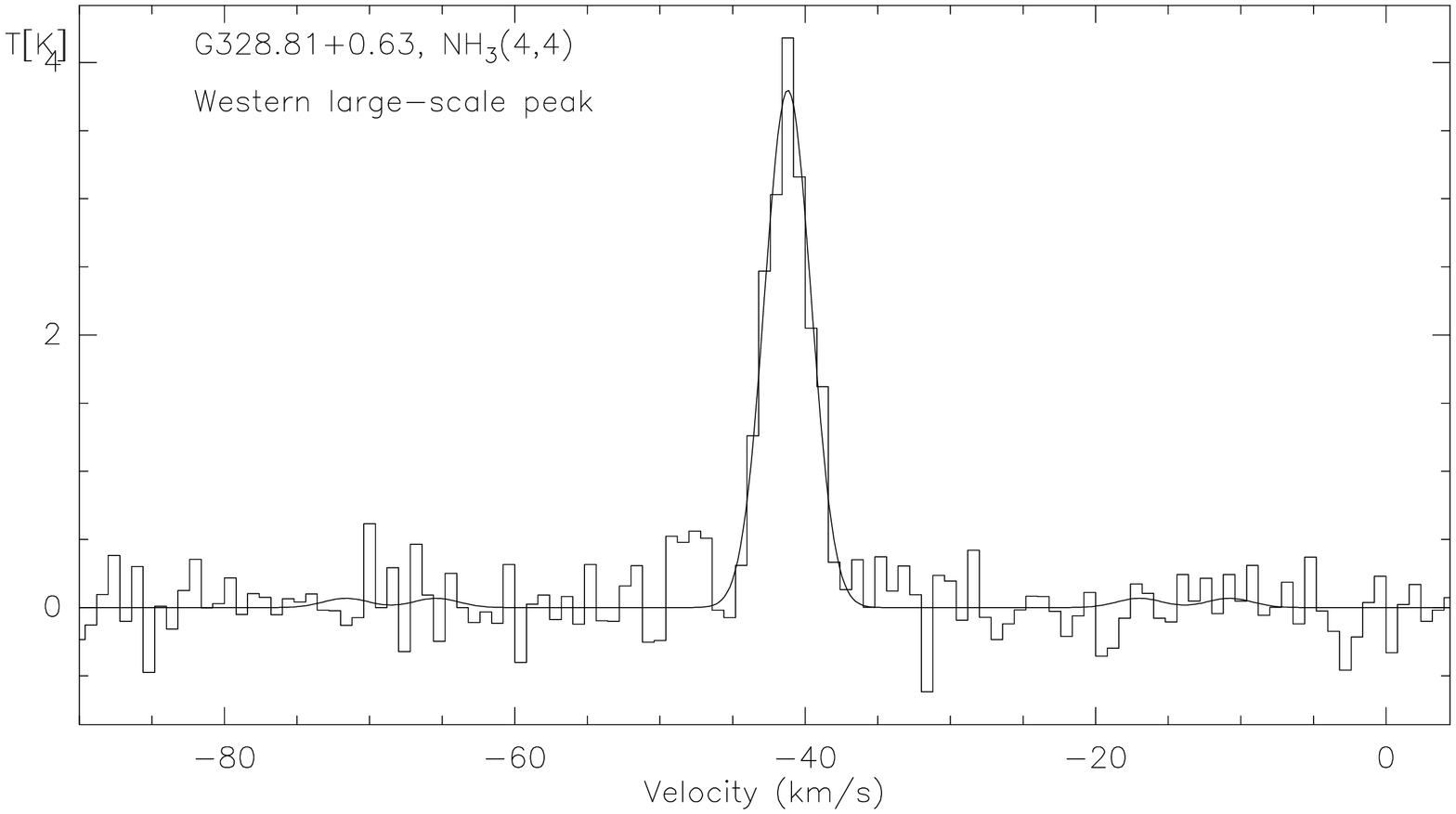}
\includegraphics[width=0.275\textwidth,angle=-90]{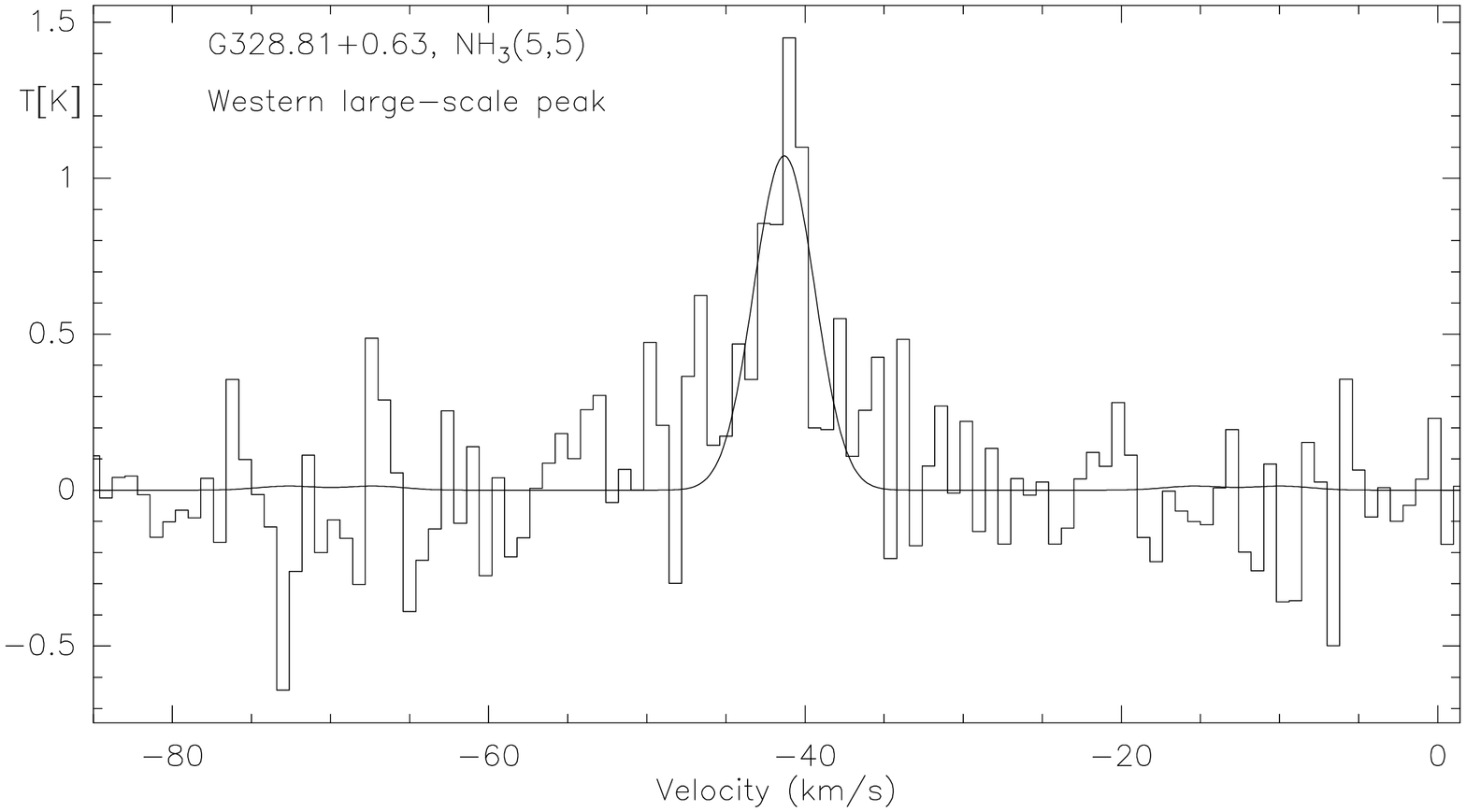}\\
\includegraphics[width=0.275\textwidth,angle=-90]{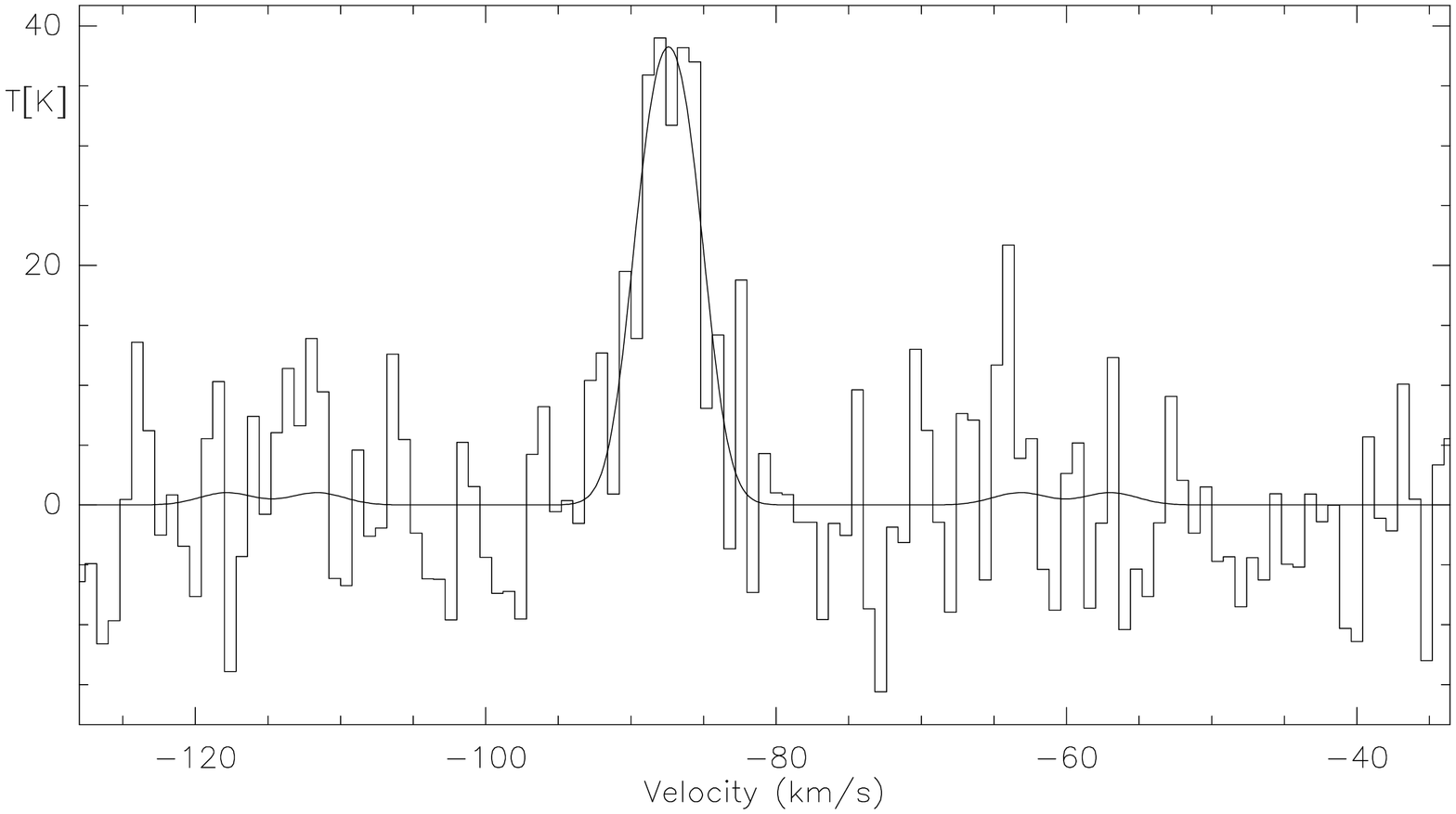}
\includegraphics[width=0.275\textwidth,angle=-90]{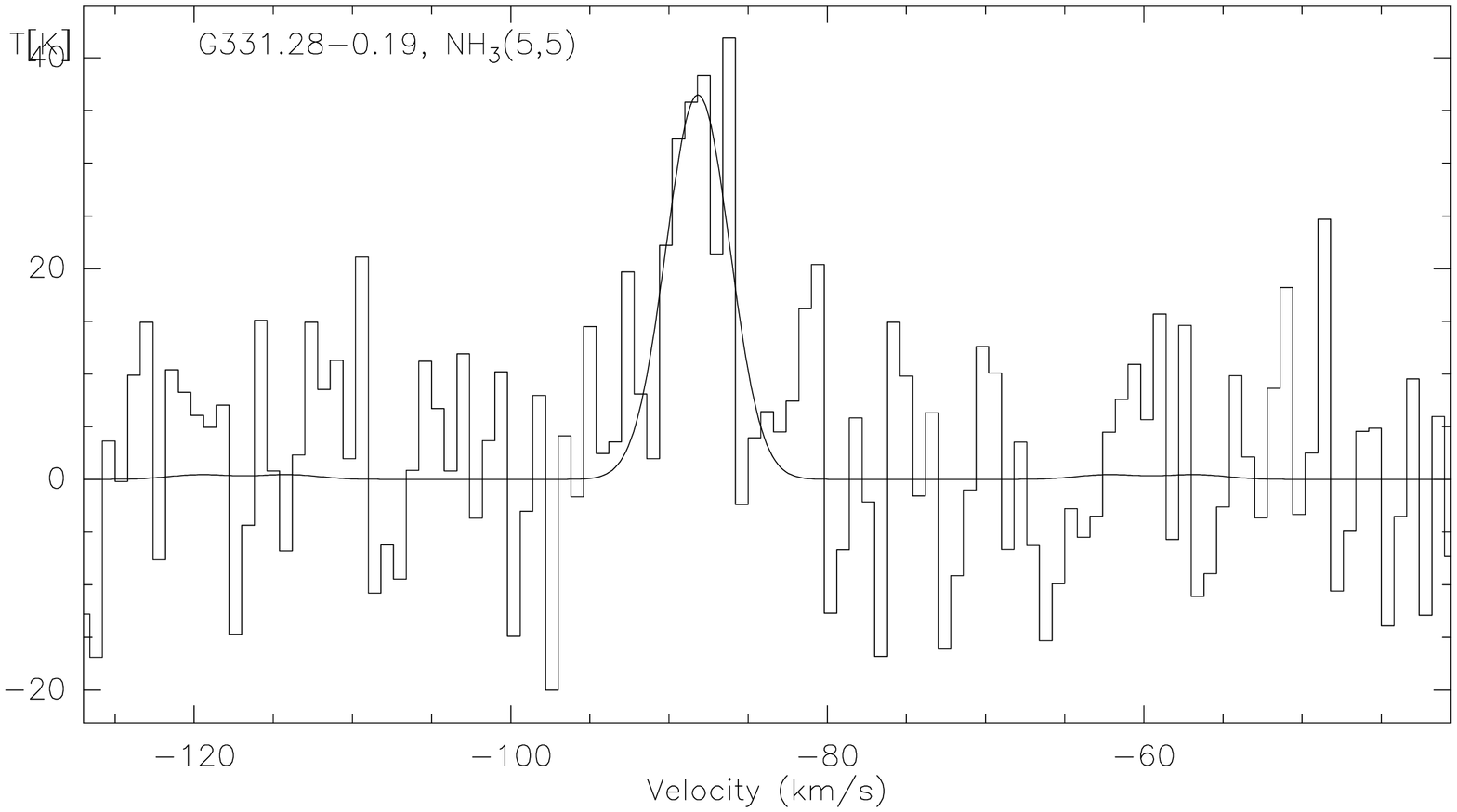}\\
\includegraphics[width=0.275\textwidth,angle=-90]{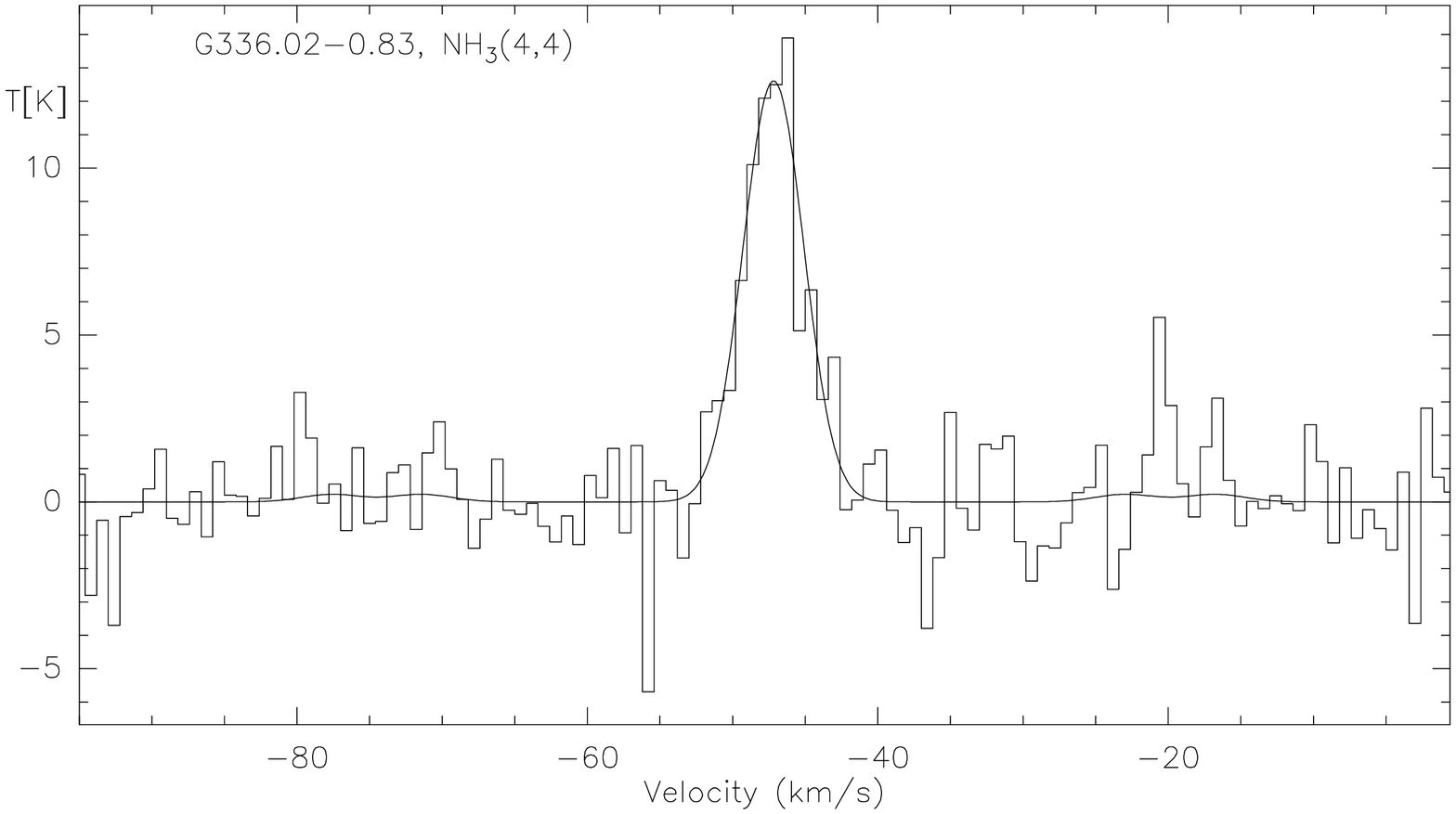}
\includegraphics[width=0.275\textwidth,angle=-90]{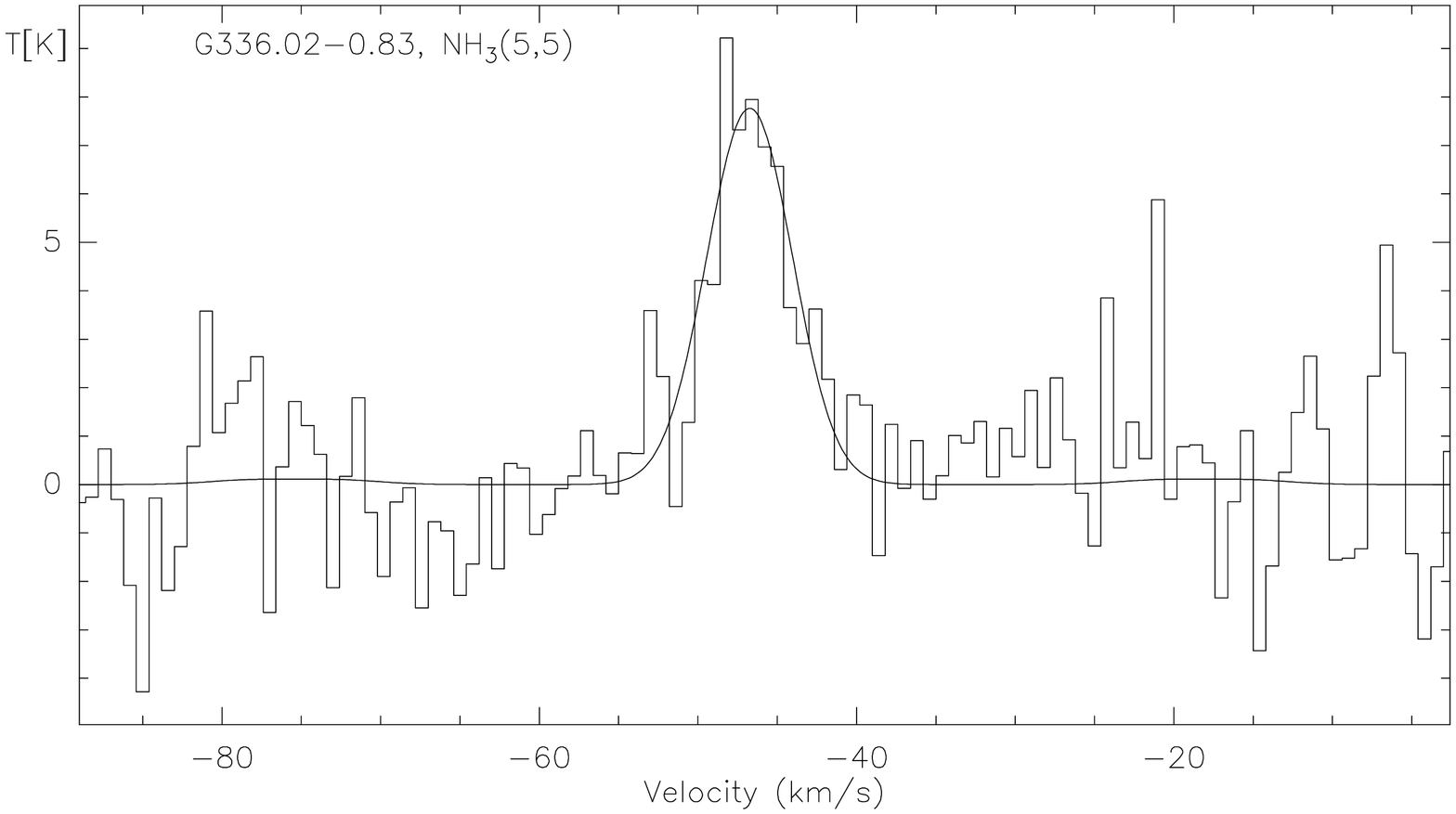}\\
\includegraphics[width=0.275\textwidth,angle=-90]{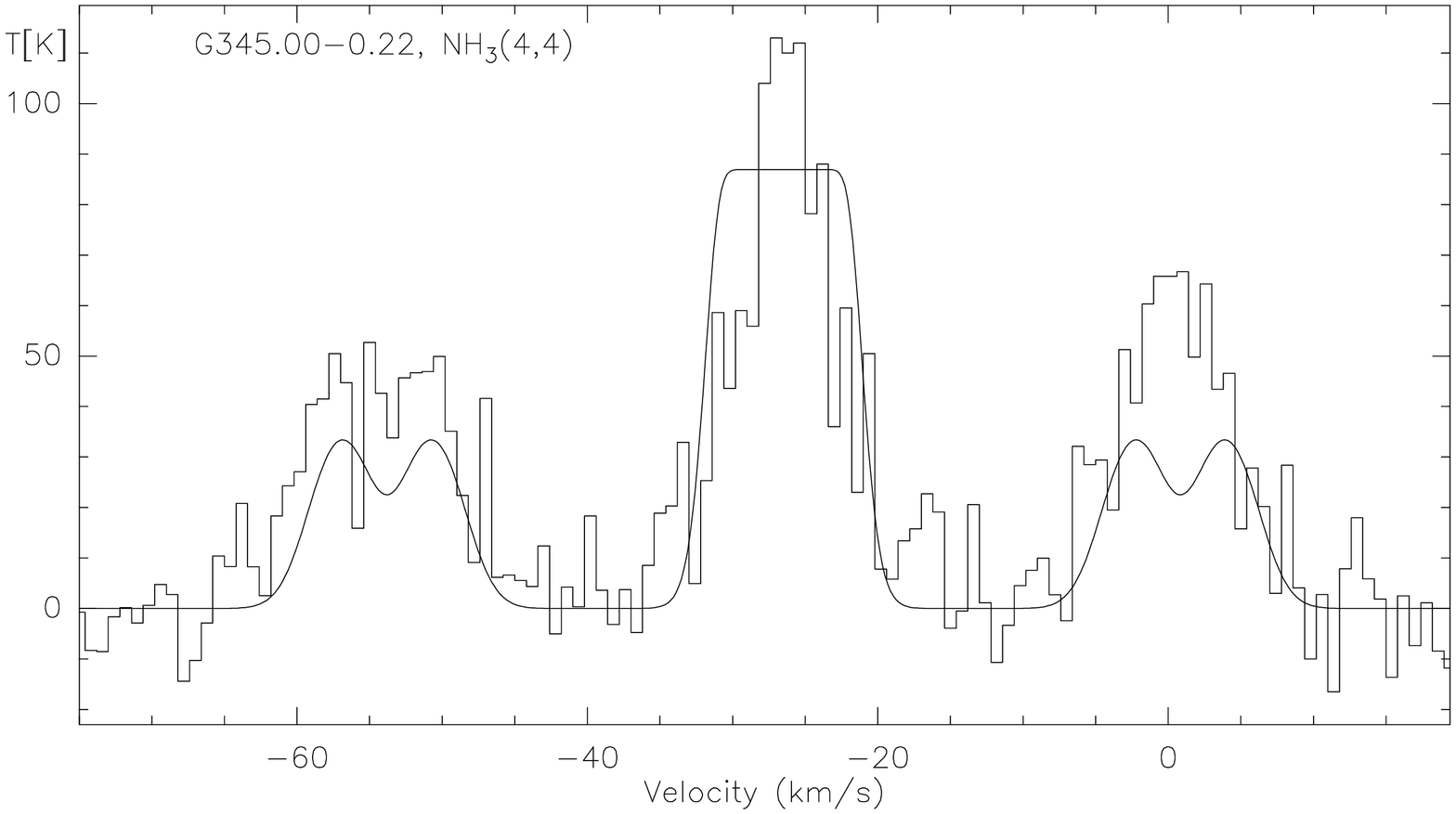}
\includegraphics[width=0.275\textwidth,angle=-90]{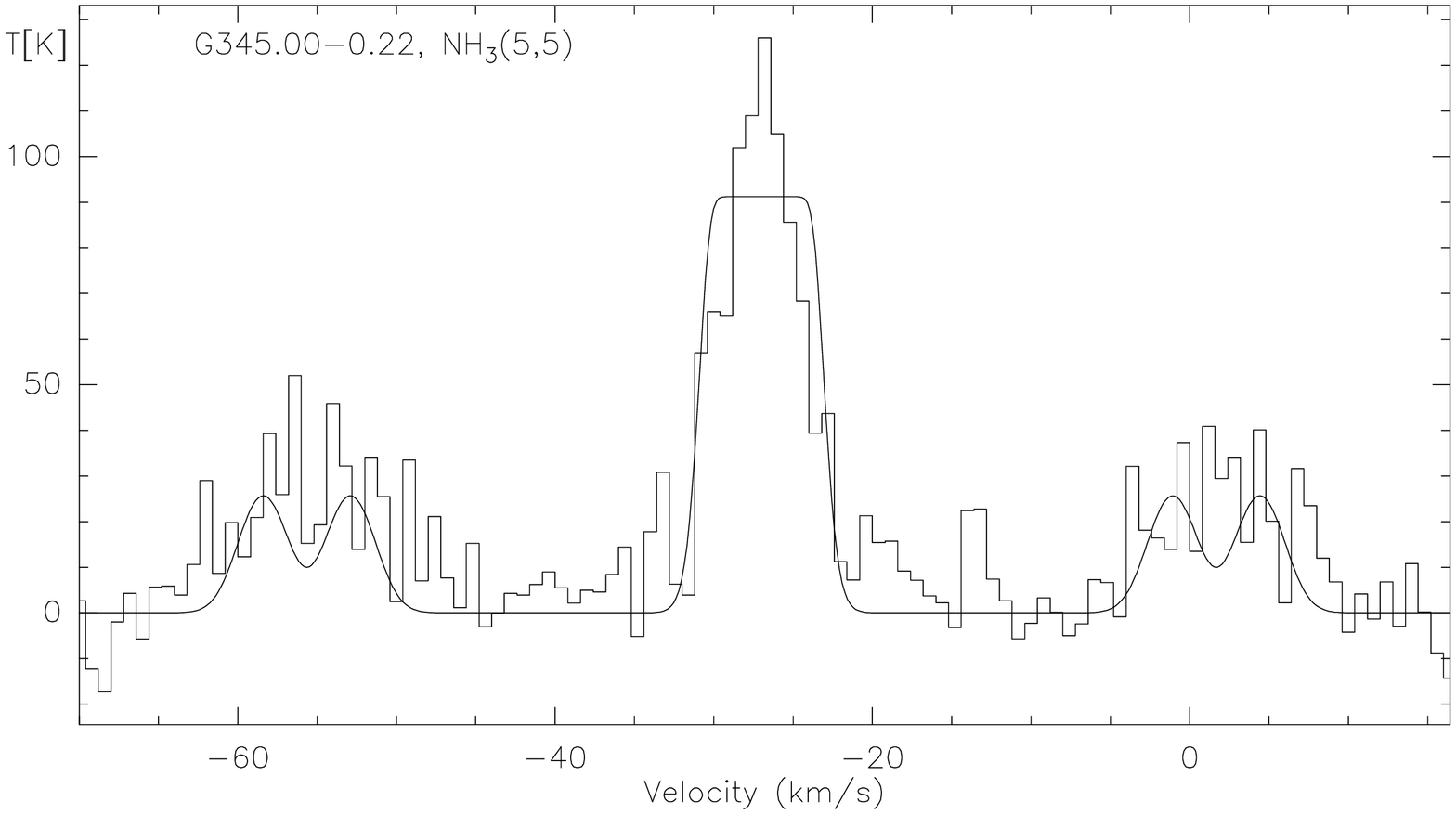}
\caption{NH$_3$(4,4) and NH$_3$(5,5) spectra (left and right column)
  extracted toward the peak positions of the sources labeled in each
  panel. For G328.81+0.63, the shown spectrum is extracted toward the
  large-scale western peak in Figure \ref{g328_nh3_large}. The histogram
  presents the data whereas the full lines show attempts to fit the
  whole hyperfine structure.  Due to the very high optical depth, even
  this hyperfine structure fitting does not work well.}
\label{spectra2}
\end{figure*}

\begin{figure*} 
\includegraphics[width=0.275\textwidth,angle=-90]{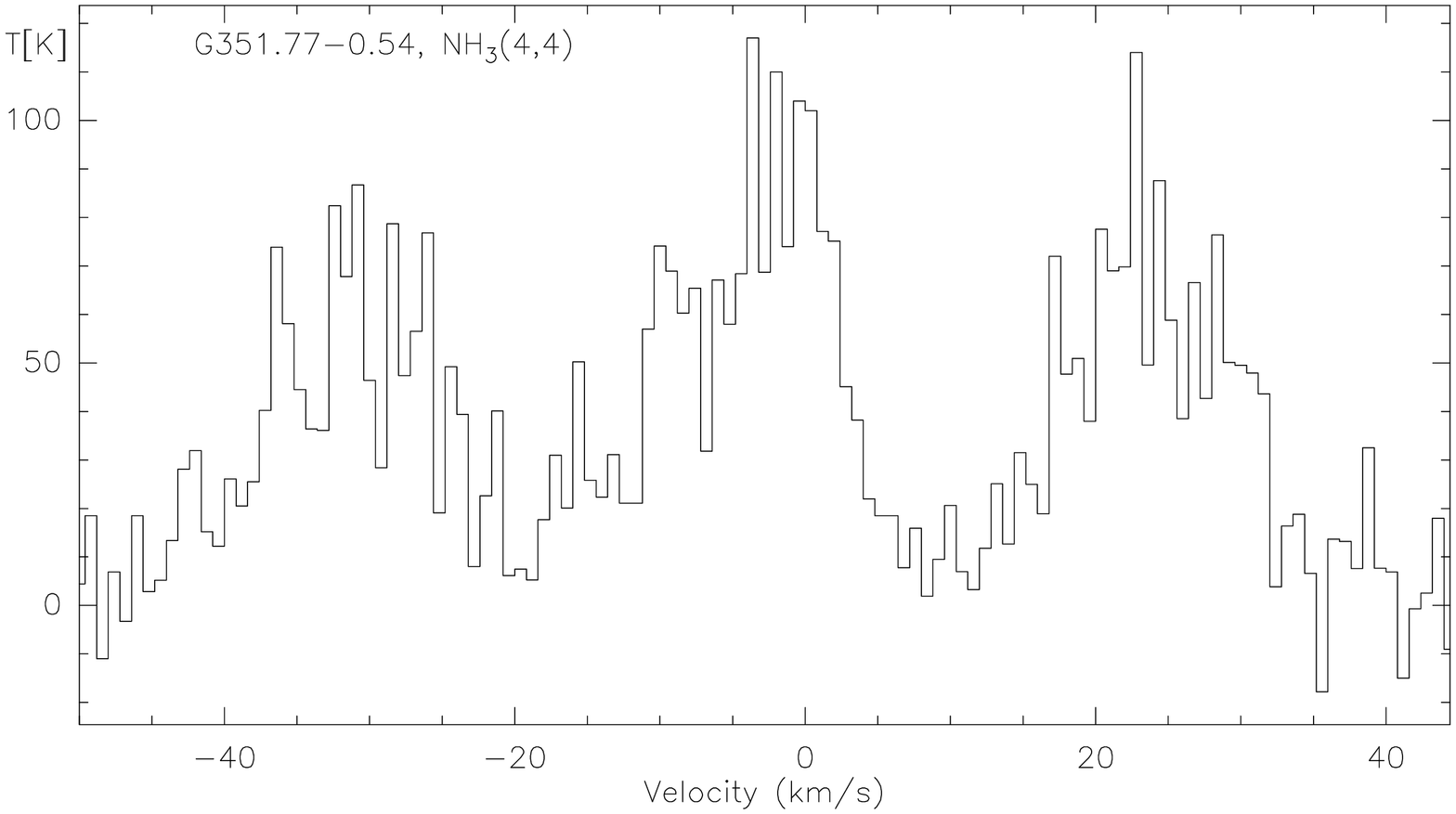}
\includegraphics[width=0.275\textwidth,angle=-90]{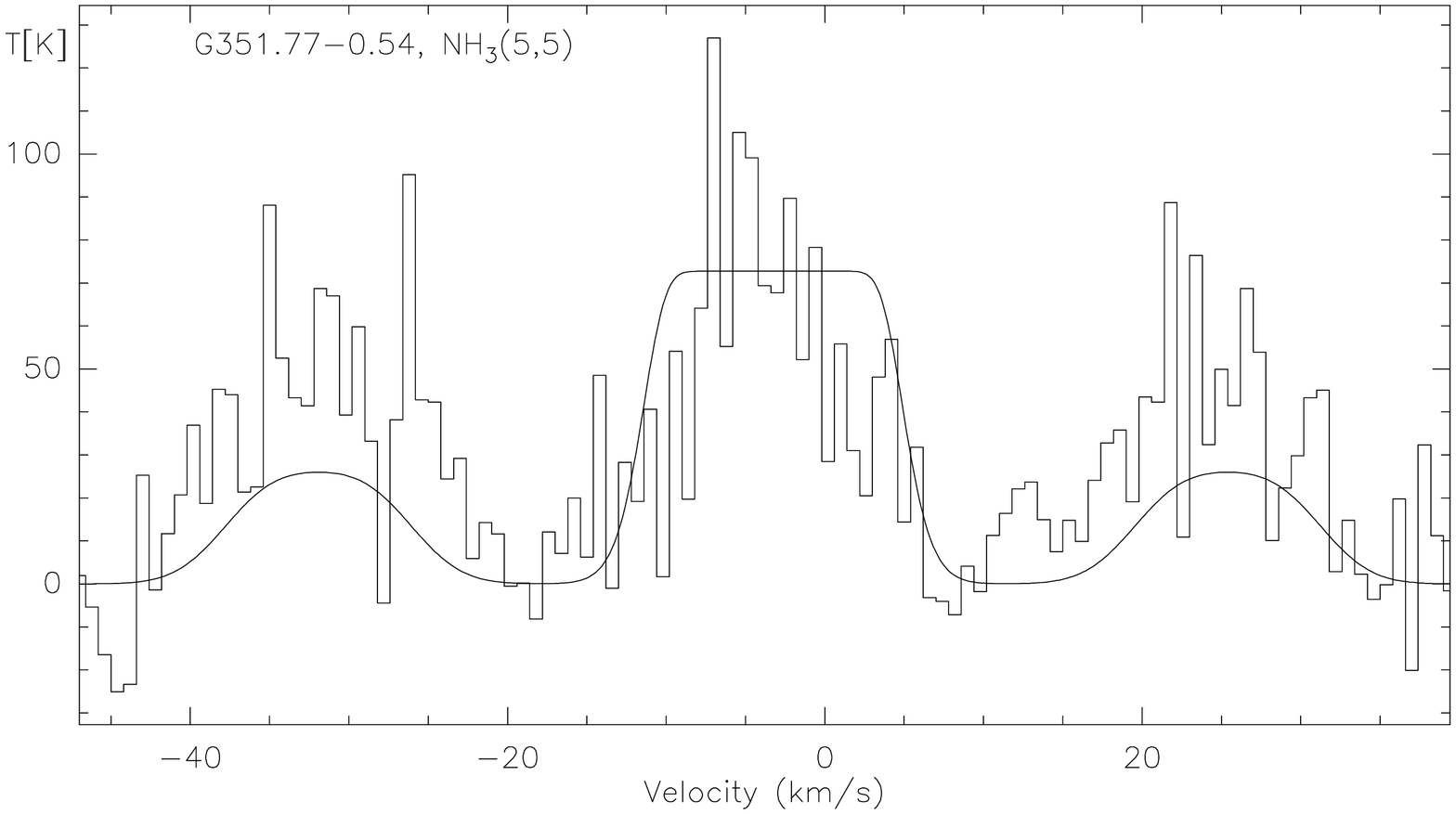}\\
\includegraphics[width=0.275\textwidth,angle=-90]{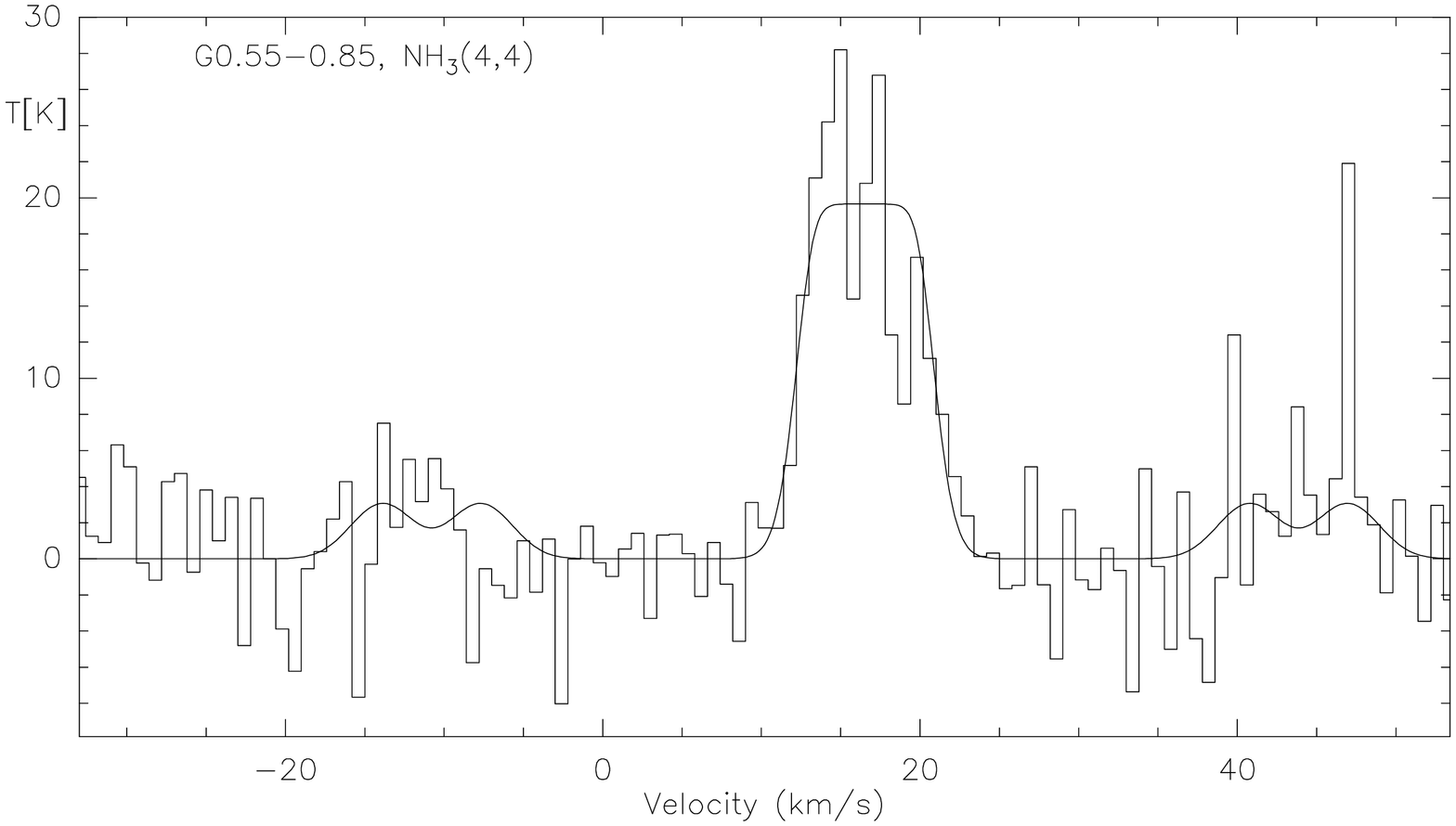}
\includegraphics[width=0.275\textwidth,angle=-90]{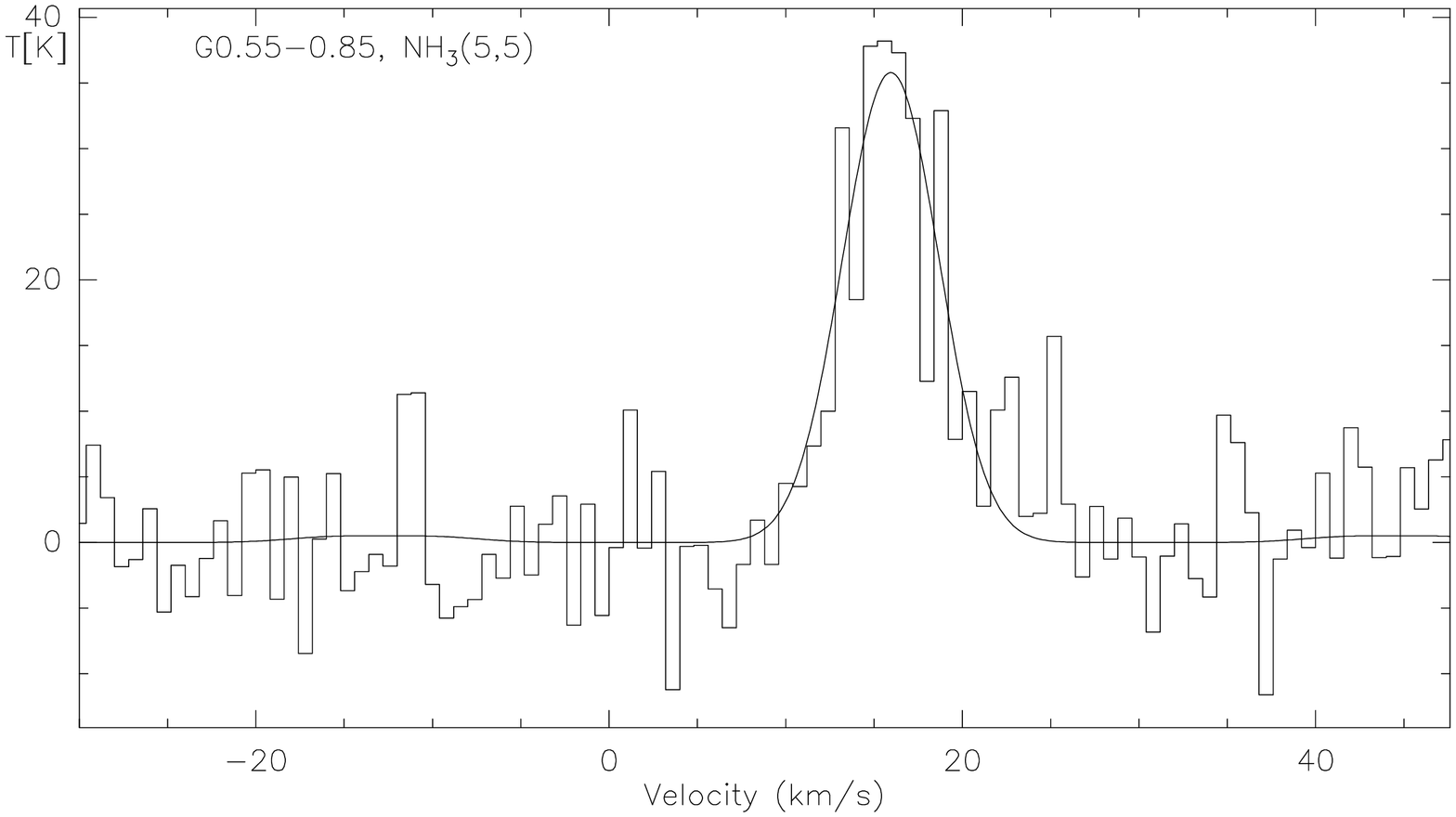}\\
\includegraphics[width=0.275\textwidth,angle=-90]{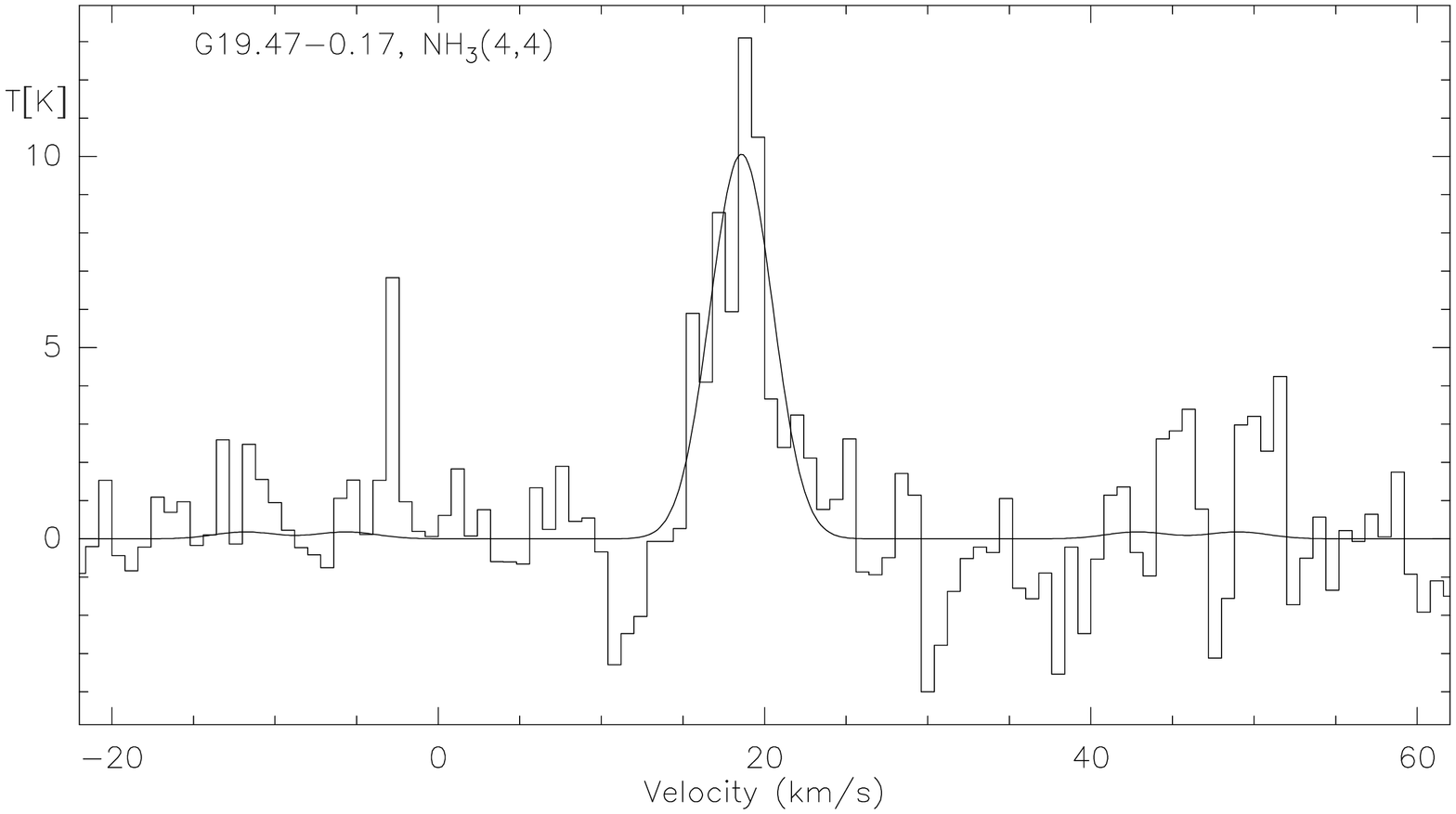}
\includegraphics[width=0.275\textwidth,angle=-90]{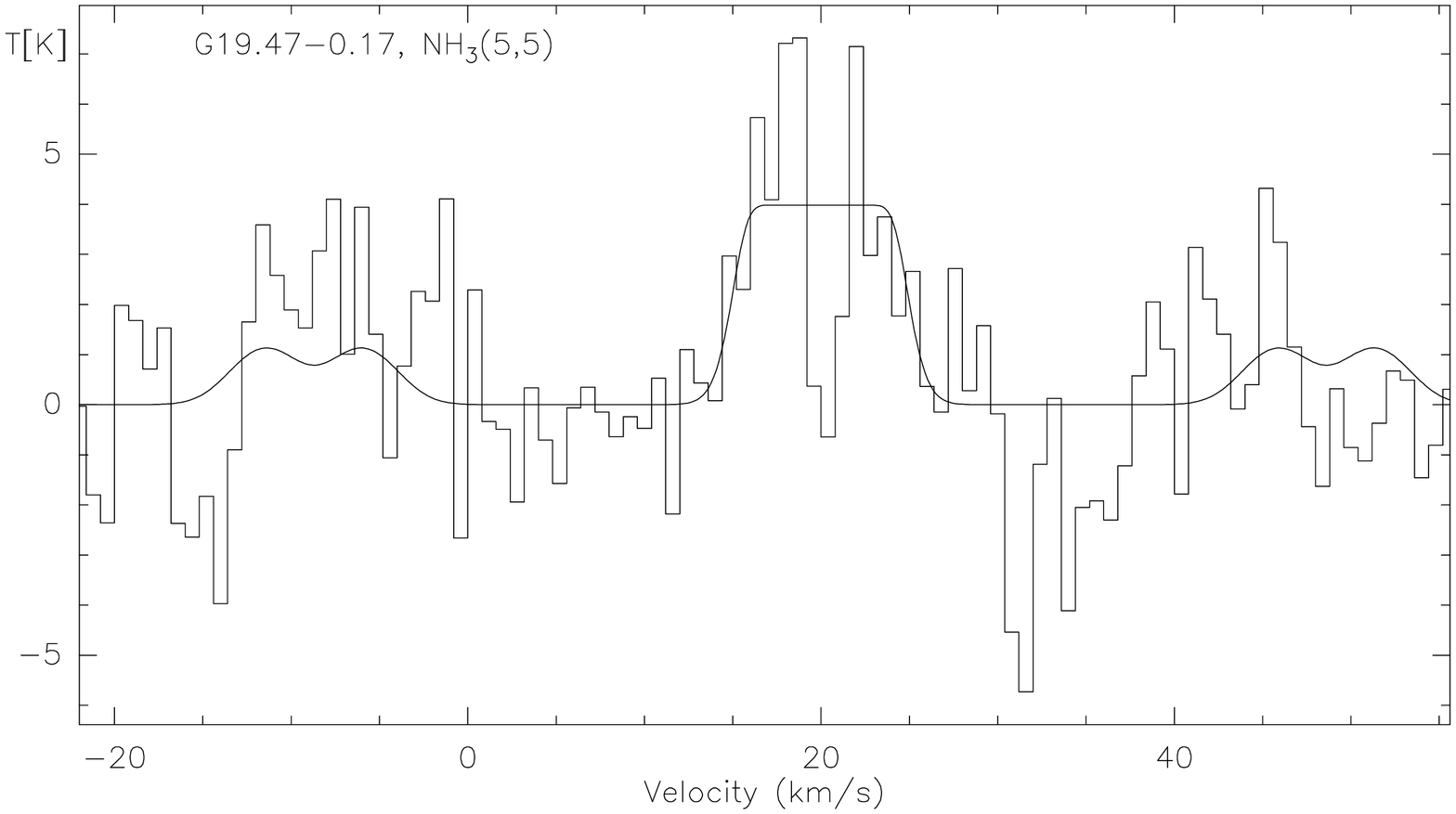}\\
\caption{NH$_3$(4,4) and NH$_3$(5,5) spectra (left and right column)
  extracted toward the peak positions of the sources labeled in each
  panel. The histogram presents the data whereas the full lines show
  attempts to fit the whole hyperfine structure.  Due to the very high
  optical depth, even this hyperfine structure fitting does not work
  well (it did not work at all for the (4,4) line of G351.77-0.54).}
\label{spectra3}
\end{figure*}

The low transition NH$_3$ lines (e.g., the (1,1) and (2,2) lines) are
known to be an excellent thermometer for the cold components of the
gas within molecular clouds (e.g., \citealt{walmsley1983}). Similarly,
we may be able to use the high-transition lines here to estimate the
temperatures of the warm gas observed in these regions. The advantage
of NH$_3$ is that one observes the whole hyperfine structure
simultaneously and hence should be able to derive the optical depth of
the lines. In the case of the NH$_3$(4,4) and (5,5) transitions, the
relative intensities in the optically thin regime of the satellite
lines with respect to the main central hyperfine components are
approximately 2 and 1\%, respectively. Figures \ref{spectra1},
\ref{spectra2} and \ref{spectra3} show example spectra of each source
and our attempts to fit the whole hyperfine structure to derive the
spectral parameters. However, in most cases the fits do not represent
the data well. This is mainly due to the extremely high optical depth
where the satellite lines reach about the same intensities as the main
central component. In such a case the fits give flat-topped spectra
which are not observed. This discrepancy indicates that there is a
temperature gradient along the line of sight which is not taken into
account by the fitting procedure. More advanced radiative transfer
calculations would be required to reproduce the spectral shape which
is out of the scope of this paper.  Therefore, we are not able to get
accurate temperature estimates for the target sources. However, based
on the high excitation temperatures of the two lines ($E_{\rm{lower}}$
of 200 and 295\,K, respectively) and the high observed brightness
temperatures between several 10 and more than 100\,K in most sources
(see Figs.~\ref{spectra1}, \ref{spectra2} and \ref{spectra3}), it is
reasonable to assume that the average gas temperatures in the observed
regions exceeds 100\,K.

While temperature gradients increasing toward the central protostars
do also increase the thermal line width toward the center, this is
unlikely to explain the central line widths increases as observed in
several of the 2nd moment maps. The thermal line width scales with the
square-root of the temperature ($\Delta v_{\rm{therm}}\propto
T^{0.5}$). Assuming a temperature gradient $T(r)\propto r^{-0.4}$, the
thermal line width scales like $\Delta v_{\rm{therm}}\propto
r^{-0.2}$. For example, assuming 100\,K temperature at a core edge
with a radius of 5000\,AU the thermal line with $\Delta
v_{\rm{therm}}$ is $\sim 0.5$\,km\,$^{-1}$. With the above relation
$\Delta v_{\rm{therm}}$ at an inner radius of 500\,AU should be
$\sim$0.8\,km\,s$^{-1}$. If we take G327.3-0.60 in
Fig.~\ref{g327_mom1} as an example, the central values exceeding
4\,km\,s$^{-1}$ are far broader than any thermal line broadening could
produce.  The associated velocity gradients suggest that the
dominating reason for the observed line width broadening should be due
to rotation.

\subsection{Morphologies of extended continuum emission}

Three of the regions we have imaged show clear evidence for extended
continuum emission -- G305.21+0.21, G316.81-0.06 and G336.02-0.83.  In
all cases, the continuum emission appears to resemble a simple
circular shape of diameter $30-40''$. We believe that in each case,
extended continuum emission is present, however, we question the
validity of the morphologies shown in the continuum images. This is
because an angular size of $\sim 100''$ corresponds to a baseline
length of approximately 2.5\,k$\lambda$, which closely matches the
shortest baseline used in these observations (31\,m). The second
shortest baseline of 199\,m corresponds to spatial scales of $\sim
15''$. While scales below $15''$ are sampled relatively well by our
observations, only a single baseline covers larger spatial scales, and
structures above $\sim 100''$ are completely filtered out.  Therefore,
our observations are sensitive to extended structures over only a
small range of sizes and thus do not represent the true morphology of
extended emission.

\begin{figure}
\includegraphics[width=0.4\textwidth,angle=-90]{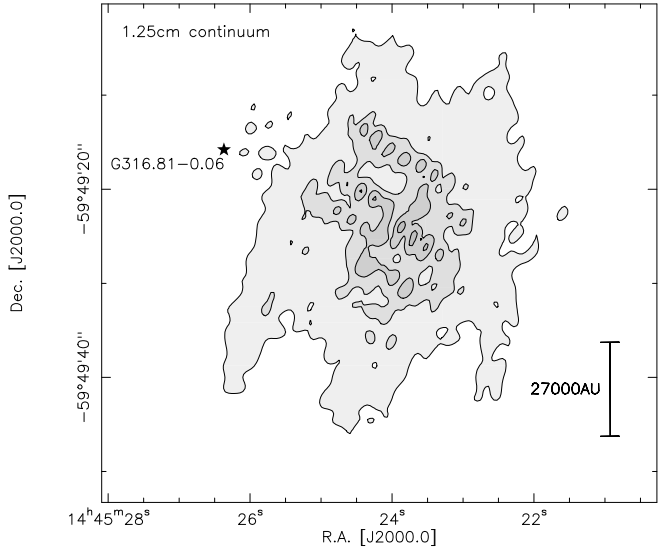}
\caption{1.25\,cm continuum emission in the field of G316.81-0.06. The
  contours start at the $5\sigma$ level and continue in $3\sigma$
  steps (Table \ref{tablesample2}). The star marks the position of our
  primary NH$_3$ target within the field, and a scale-bar is shown at
  the bottom-right.}
\label{g316_cont}
\end{figure}

Recent observations by \citet{longmore2009} of G316.81-0.06 at
18.8\,GHz show two sources on either side of the emission we show in
Figure \ref{g316_cont}.  These recent observations include a better
sampling of the UV plane over scales corresponding to the extent of
the emission found by \citet{longmore2009} and so are a more accurate
representation of the emission in this region. To double-check, by
excluding baselines $>4$\,k$\lambda$ from the continuum data presented
in \citet{longmore2009}, we recover an image similar to that in the
current paper.

We therefore conclude that the morphology of continuum emission in
G305.21+0.21, G316.81--0.06 and G336.02--0.83 is probably not
accurately represented by that shown in Figures \ref{g305_cont},
\ref{g316_cont} and \ref{g336_cont}, respectively. We caution the
reader on interpretation of extended emission with similar
interferometer configurations.

\section{General implications}
\label{general}

Table \ref{results} summarizes the general results regarding
rotational signatures for the whole sample. Except for the source
IRAS\,18151-1208 which did not satisfy the original selection criteria
from the rest of the sample, all other sources were clearly detected
in the high-excitation NH$_3$ lines, implying that our sample
selection criteria were well chosen. Out of the remaining eleven
sources, six show signatures of rotation and/or infall which can be
considered as strong evidence of ongoing high-mass star formation
activity. While the rotational signatures vary between typical
velocity gradients perpendicular to the outflows to more spherical
infall signatures and infall signatures from absorption lines, we do
not find clear signs of Keplerian rotation. This implies that although
we have achieved very high angular resolution, mostly better than
$1''$ (Table \ref{tablesample2}), the corresponding linear scales
(Tables \ref{tablesample} \& \ref{tablesample2}) do not follow typical
disk-like Keplerian signatures. Since outflows and jets are known for
most of the sources, and since these are believed to be accelerated
via disk-winds (e.g., \citealt{arce2006}), the corresponding accretion
disks should be smaller. Furthermore, the spectra presented in
Figs.~\ref{spectra1}, \ref{spectra2} and \ref{spectra3} show that even
highly excited lines still have a very high optical depth. Hence we
are only seeing the $\tau =1$ surface of these spectral lines and do
not trace the innermost regions. Hence the accretion disks can still
be obscured by the high optical depths.

Compared to other examples where spatial flattening is observed on
scales $> 10^4$\,AU (e.g., M17 \citealt{chini2004}, IRDC18223-3
Fallscheer et al.~2009), it is surprising that we do not find a
single source where spatial flattening is observed in our data. To
the contrary, in several sources where we see clear velocity gradients
indicative of rotation, the spatial structure of the gas is
distributed over scales $> 10^4$\,AU in a very clumpy fashion (e.g.,
G327.3-0.60 or G351.77-0.54). While we find infall signatures in
absorption lines (e.g., G328.81+0.63), it is also interesting to note
that in at least one source (G327.3-0.60) we observe the ``bull's-eye''
signature in the 1st moment map that was attributed to spherical
infall motions by \citet{sollins2005} for one of the most high-mass
star-forming regions G10.6-0.4.

The high-excitation lines are always found toward the CH$_3$OH Class
II maser positions, reinforcing the idea that these masers are
excellent tracers for early stages of high-mass star formation.
However, the data do not allow us to draw clear conclusions whether
these maser are disk or outflow associated.

In addition to the NH$_3$ data, toward most fields we detect 1.25\,cm
continuum emission from either UCH{\sc ii} regions associated with our
targets, or, more typically, extended continuum emission offset more
than $10''$ tracing more evolved H{\sc ii} regions in the field of
view, which however is not directly associated with our target
sources. This highlights that high-mass star formation rarely proceeds
in isolation but that other potentially more evolved regions often can
exist within the same star formation complex..

\begin{table*}[htb]
\centering
\caption{Summary results for rotational signatures.}
\begin{tabular}{ll}
\hline \hline
Source & Results \\
\hline
G305.21+0.21 & Rotating structure perpendicular to outflow, no Keplerian signature. \\
G316.81-0.06 & Inconclusive.\\
G323.74-0.26 & Inconclusive.\\
G327.3-0.60  & Infall and rotational signatures. Clumpy large structure.\\
G328.81+0.63 & Infall toward UCH{\sc ii} regions. Large-scale hot gas in emission.\\
G331.28-0.19 & Velocity gradient perpendicular to outflow suggestive of rotation and infall.\\
             &  However, differences between lines. \\
G336.02-0.83 & Inconclusive.\\
G345.00-0.22 & Rotation in western peak, expanding UCH{\sc ii} region in eastern peak.\\
G351.77-0.54 & Rotating structure perpendicular to outflow, no Keplerian signature. \\
G0.55-0.85   & Inconclusive.\\
G19.47-0.17  & Inconclusive.\\
I18151-1208  & Non-detection\\
\hline
\end{tabular}
\label{results}
\end{table*}

\section{Conclusion}

We present a large observing campaign to search for rotational motions
in high-mass star-forming regions via high-excitation NH$_3$ line
observation with high spatial resolution. While we detect all sources
from the original sample (excluding the one source that was chosen via
other selection criteria), in more than 50\% of them rotational and/or
infall motions can be identified. Several more general conclusions
can be drawn from this observing campaign.\\

\noindent (a) High-excitation NH$_3$ lines are very good tracers of
the dense
gas within hot-core type young high-mass star-forming regions. In this sample, the NH$_3$(4,4) and (5,5) emission is always associated with the CH$_3$OH Class II methanol maser emission, which is a well-known signpost for high-mass star formation.\\
(b) We identify rotational/infalling motions in half of the observed
sources.\\
(c) The signatures comprise simple velocity gradients as well as more
spherical infall signatures.\\
(d) Although the spatial resolution is of order a few 1000\,AU, we do not
identify obvious Keplerian signatures.\\
(e) While high optical depth is an issue, the data nevertheless
indicate that accretion disks resembling those of low-mass
star-forming regions (e.g., \citealt{simon2000}) have to be typically
of smaller sizes, not yet resolved by our observations. Likely such
smaller accretion disks
could be fed by the rotating envelopes observed here.\\
(f) In many fields of view we do detect additional H{\sc ii} regions
that are not directly linked to our targets but that reside in the
same projected area on the sky.  This is additional evidence for the
clustered mode of high-mass star formation.\\

Where to go from here? Clearly there are different paths to be
followed. While our approach for this study was biased toward hot-core
type sources based on our spectral line selection criteria, other
observational approaches can target sources in different evolutionary
stages as well as of different luminosities. In particular,
observations at mm wavelengths with broad spectral bandpasses allow us
to observe several spectral lines simultaneously which can trace
rotational motions in sources of different characteristics (e.g.,
\citealt{zhang2005b,cesaroni2006,beuther2009a}). For example, Qizhou
Zhang and collaborators started a similarly large project searching
for disk signatures toward a sample of young high-mass star-forming
regions with the Submillimeter Array, and they also detected
rotational signatures toward all targets (Qizhou Zhang et al., in
prep.). While as outlined in section \ref{general} flattened
structures do exist on scales exceeding $10^4$\,AU, most data indicate
that the Keplerian accretion disks should reside on smaller scales,
probably $\leq 1000$\,AU. This is also shown in the recent
3-dimensional radiative hydrodynamic simulations by
\citet{krumholz2006b,krumholz2009} where most of the disk-like
structures are contained with a radius of $\sim$500\,AU. Furthermore,
recent work has also shown that accretion disks exceeding 150\,AU are
prone to fragmentation (e.g., \citealt{kratter2006,vaidya2009}), hence, it is not even expected to find Keplerian signatures on
much larger scales.

While we can already achieve tremendous progress with the existing
instrumentation, clearly we are still lacking spatial resolution if we
want to understand the structure and physical processes of the
more proper central accretion disks. Future instruments like ALMA and
eVLA will allow us to make considerable progress in this direction for
the cold gas components. Furthermore, due to the large accretion
rates, strong viscous forces and central luminous sources, a
considerable fraction of warm/hot gas is expected to exist with the
high-mass disks. Therefore, imaging these warm/hot dust and gas
components at high spatial resolution with future instruments like
MIRI on JWST will allow us to constrain the properties of these
components in much more detail. It is likely that only a concerted
effort at long and short wavelengths combined with hydrodynamic and
radiation transfer modeling will give us a real understanding of the
working of high-mass accretion disks.




\acknowledgments
  The Australia Telescope Compact Array is part of the Australia
  Telescope which is funded by the Commonwealth of Australia for
  operation as a National Facility managed by CSIRO.
  H.B.~acknowledges financial support by the Emmy-Noether-Program of
  the Deutsche Forschungsgemeinschaft (DFG, grant BE2578).

\begin{figure*}
\centering
\caption{G305.21+0.21 intensity weighted velocity (1st moment, top
  row) and line width (2nd moment, bottom row) maps of the main
  hyperfine components of NH$_3$(4,4) (left) and NH$_3$(5,5) (right).
  The triangle marks the main Class II CH$_3$OH maser position from
  \citet{norris1993}, the arrows in the top-left panel outline the
  approximate direction from the assumed outflow (e.g.,
  \citealt{debuizer2003}), the synthesized beams are shown at the
  bottom-left of each panel, a scale-bar is presented in the top-left
  panel, and the 0/0 position is given in Table \ref{tablesample}.}
\label{g305_mom1}
\end{figure*}

\begin{figure*}
\centering
\caption{G316.81-0.06 intensity weighted velocity (1st moment, top
  row) and line width (2nd moment, bottom row) maps of the main
  hyperfine components of NH$_3$(4,4) (left) and NH$_3$(5,5) (right).
  The triangles mark the Class II CH$_3$OH maser positions
  \citep{walsh1998}, and the synthesized beams are shown at the
  bottom-left of each panel, a scale-bar is presented in the top-left
  panel, and the 0/0 position is given in Table \ref{tablesample}.}
\label{g316_mom1}
\end{figure*}

\begin{figure*}
\centering
\caption{G323.74-0.26 intensity weighted velocity (1st moment, top
  row) and line width (2nd moment, bottom row) maps of the main
  hyperfine components of NH$_3$(4,4) (left) and NH$_3$(5,5) (right).
  The triangles mark the Class II CH$_3$OH maser positions
  \citep{walsh1998}, the synthesized beams are shown at the
  bottom-left of each panel, a scale-bar is presented in the top-left
  panel, and the 0/0 position is given in Table \ref{tablesample}.}
\label{g323_mom1}
\end{figure*}


\begin{figure*}
\centering
\caption{G327.3-0.60 intensity weighted velocity (1st moment) maps of
  the main hyperfine components of NH$_3$(4,4) (left) and NH$_3$(5,5)
  (right) excluding antenna 6 to better show the larger-scale
  structure. The triangles mark the positions of the Class II CH$_3$OH
  masers (James Caswell, private communication as referred to by
  \citealt{caswell1998}), the synthesized beams are shown at the
  bottom-right of each panel, a scale-bar assuming near kinematic
  distance is presented in the left panel, and the 0/0 position is
  given in Table \ref{tablesample}.}
\label{g327_mom1_noant6}
\end{figure*}

\begin{figure*}
\centering
\caption{G327.3-0.60 intensity weighted velocity (1st moment, top row)
  and line width (2nd moment, bottom row) maps of the main hyperfine
  components of NH$_3$(4,4) (left) and NH$_3$(5,5) (middle) including
  also antenna 6 data to also show the smaller-scale sub-structure.
  The right panel shows the 1st moment map of the most blue-shifted
  NH$_3$(4,4) hyperfine component. The triangles mark the positions of
  the Class II CH$_3$OH masers (James Caswell, private communication
  as referred to by \citealt{caswell1998}), the synthesized beams are
  shown at the bottom-left of each panel, a scale-bar assuming near
  kinematic distance is presented in the bottom-left panel, and the
  0/0 position is given in Table \ref{tablesample}.}
\label{g327_mom1}
\end{figure*}

\clearpage

\begin{figure}
\includegraphics[width=0.43\textwidth,angle=-90]{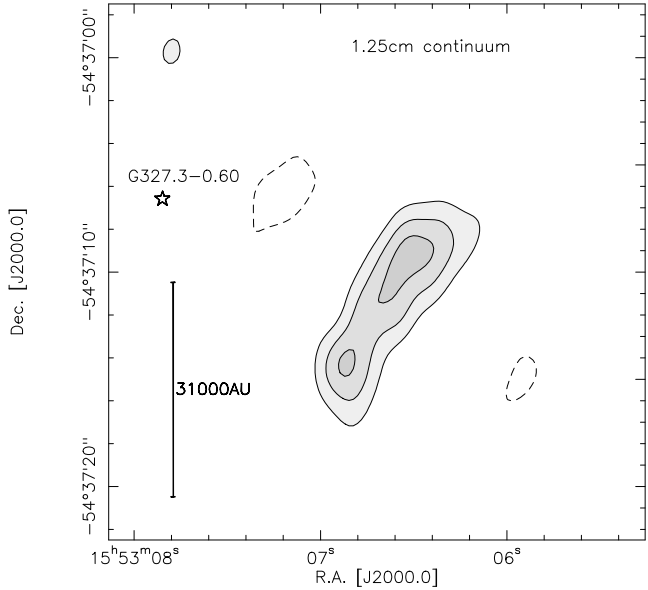}
\caption{1.25\,cm continuum emission in the field of G327.3-0.60. The
  contour levels start at $3\sigma$ and continue in $2\sigma$ steps
  (Table \ref{tablesample2}, full lines positive, and dashed lines
  negative features).  The star marks the position of our primary
  NH$_3$ target within the field, and a scale-bar is shown at
  the bottom-right.}
\label{g327_cont}
\end{figure}

\begin{figure}
\caption{G328.81+0.63: The grey-scales present the 1st moment maps of
  the absorption features of the main NH$_3$(4,4) and (5,5) hyperfine
  components (top and bottom panels, respectively) against the
  ultracompact H{\sc ii} region outlined in contours from the 1.25\,cm
  continuum emission. The contouring is done in $6\sigma$ steps of
  16.8\,mJy\,beam$^{-1}$. The triangles mark the Class II CH$_3$OH
  maser positions \citep{walsh1998}, and the synthesized beams and
  scale-bars are plotted at the bottom left of each panel.}
\label{g328_mom1}
\end{figure}

\begin{figure}
\caption{G328.81+0.63 centered on the brightest eastern peak from Figure
  \ref{g328_mom1}: The grey-scale presents the 1st moment maps of the
  absorption features of the main NH$_3$(4,4) hyperfine component
  against the ultracompact H{\sc ii} region outlined in contours from
  the 1.25\,cm continuum emission. The contouring is done in $6\sigma$
  steps of 16.8\,mJy\,beam$^{-1}$.}
\label{g328_mom1_peaks}
\end{figure}

\begin{figure}
\caption{G328.81+0.63: The grey-scales present the 2nd moment maps of
  the absorption features of the main NH$_3$(4,4) hyperfine component
  against the ultracompact H{\sc ii} region outlined in contours from
  the 1.25\,cm continuum emission. The contouring is done in $6\sigma$
  steps of 16.8\,mJy\,beam$^{-1}$. The triangles mark the Class II
  CH$_3$OH maser positions \citep{walsh1998}, and the synthesized beam
  and scale-bar are plotted at the bottom left of each panel.}
\label{g328_mom2}
\end{figure}

\clearpage

\begin{figure*}
\centering
\caption{G331.28-0.19 intensity weighted velocity (1st moment, top
  row) and line width (bottom row) maps of the main hyperfine
  components of NH$_3$(4,4) (left) and NH$_3$(5,5) (right). The
  contours show the 1.25\,cm continuum emission with contour levels in
  $3\sigma$ steps of 0.9\,mJy\,beam$^{-1}$. The white triangles mark
  the Class II CH$_3$OH maser positions \citep{walsh1998}, the arrows
  in the top-left panel outline the approximate direction of the
  outflow (Table \ref{tablesample}), the synthesized beams are shown
  at the bottom-left of each panel, a scale-bar is presented in the
  top-left panel, and the 0/0 position is given in Table
  \ref{tablesample}.}
\label{g331_mom1}
\end{figure*}

\begin{figure*}
\centering
\caption{G336.02-0.83 intensity weighted velocity (1st moment, top
  row) and line width (2nd moment, bottom row) maps of the main
  hyperfine components of NH$_3$(4,4) and NH$_3$(5,5) excluding the
  long baselines associated with antenna 6. The triangles mark the
  Class II CH$_3$OH maser positions by \citet{walsh1998}, the
  synthesized beams are shown at the bottom-left of each panel, a
  scale-bar adopting the near kinematic distance is presented in the
  top-left panel, and the 0/0 position is given in Table
  \ref{tablesample}.}
\label{g336_mom1}
\end{figure*}

\begin{figure*}
\centering
\caption{G345.00-0.22 intensity weighted velocity maps (1st moment) of
  the main hyperfine components of NH$_3$(4,4) and (5,5) in the left
  and right panels, respectively. The contours show the 1.25\,cm
  continuum emission starting at the $4\sigma$ level and continuing in
  $6\sigma$ steps. While the main panels show the NH$_3$ in emission,
  the inset presents the absorption against the continuum. The
  triangles mark the Class II CH$_3$OH maser positions by
  \citet{walsh1998}, the synthesized beams are shown at the
  bottom-left of the two bottom panels, a scale-bar adopting the near
  kinematic distance is presented in the top-left panel, and the 0/0
  position is given in Table \ref{tablesample}.}
\label{g345_mom1}
\end{figure*}

\begin{figure*}
\centering
\caption{G345.00-0.22 intensity line width maps (2nd moment) of the
  main hyperfine components of NH$_3$(4,4) and (5,5) in the left and
  right panels, respectively. The contours show the 1.25\,cm continuum
  emission starting at the $4\sigma$ level and continuing in $6\sigma$
  steps. The triangles mark the Class II CH$_3$OH maser positions by
  \citet{walsh1998}, the synthesized beams are shown at the
  bottom-left of the two bottom panels, a scale-bar adopting the near
  kinematic distance is presented in the top-left panel, and the 0/0
  position is given in Table \ref{tablesample}.}
\label{g345_mom2}
\end{figure*}


\begin{figure*}
\centering
\caption{G351.77-0.54 intensity weighted velocity (1st moment) maps of
  the main hyperfine components of NH$_3$(4,4) (left) and NH$_3$(5,5)
  (right).  The triangles mark the positions of the Class II CH$_3$OH
  masers \citep{walsh1998}, the arrows and stars in the left panel
  outline the approximate direction of the outflow (Table
  \ref{tablesample}) and the positions of the cm sources from
  \citet{zapata2008}.  The synthesized beams are shown at the
  bottom-left, a scale-bar is presented in the bottom-left panel, and
  the 0/0 position is given in Table \ref{tablesample}.}
\label{g351_mom1_noant6}
\end{figure*}

\begin{figure*}
\centering
\caption{G351.77-0.54 intensity weighted velocity (1st moment, top
  row) and line width (2nd moment, bottom row) maps of the main
  hyperfine components of NH$_3$(4,4) (left) and NH$_3$(5,5) (right).
  The triangles mark the positions of the Class II CH$_3$OH masers
  \citep{walsh1998}, the arrows and stars in the top-left panel
  outline the approximate direction of the outflow (Table
  \ref{tablesample}) and the positions of the cm sources from
  \citet{zapata2008}.  The synthesized beams are shown at the
  bottom-left of each panel, a scale-bar is presented in the top-left
  panel, and the 0/0 position is given in Table \ref{tablesample}.}
\label{g351_mom1}
\end{figure*}

\begin{figure*}
\centering
\caption{G0.55-0.85 intensity weighted velocity (1st moment, top row)
  and line width (2nd moment, bottom row) maps of the main hyperfine
  components of NH$_3$(4,4) and (5,5). The triangles mark the Class II
  CH$_3$OH maser positions by \citet{walsh1998}, the synthesized beams
  are shown at the bottom-right of each panel, a scale-bar adopting
  the near kinematic distance is presented in the top-left panel, and
  the 0/0 position is given in Table \ref{tablesample}.}
\label{g055_mom1}
\end{figure*}

\end{document}